\documentclass[journal]{IEEEtran}
\usepackage{textcomp}
\usepackage{xcolor}
\usepackage{graphicx}
\usepackage{amsmath}
\usepackage{amssymb}
\usepackage{amsfonts}
\usepackage{epsfig}
\usepackage{epstopdf}
\usepackage{hhline}
\usepackage{stfloats}
\usepackage{cite}
\usepackage{algorithm}
\usepackage{algorithmic}
\usepackage{multirow}
\usepackage{subfigure}
\usepackage[CJKbookmarks=true,
bookmarksnumbered=true,
bookmarksopen=true]{hyperref}
\usepackage{balance}
\usepackage{booktabs} 
\usepackage{caption}

\begin{document}
 
\title{Physics-Informed Path-Parametric Learning for Efficient and Lightweight CSI Feedback}

\author{
Chunyu~Ling, 
Jiajia~Guo, \emph{Member}, \emph{IEEE},
Yiming~Cui,
Chao-Kai~Wen, \emph{Fellow}, \emph{IEEE}, 
Shi~Jin, \emph{Fellow}, \emph{IEEE},

Shuangfeng~Han, \emph{Senior Member}, \emph{IEEE} and
Xiaoyun~Wang



	


\thanks{An earlier version of this paper was accepted by IEEE ICC 2026 \cite{icc}.}

}		

\maketitle

\begin{abstract}

Channel State Information (CSI) feedback is vital for high spectral efficiency in wireless systems, yet high-dimensional CSI introduce significant feedback overhead. Recent deep learning (DL) approaches alleviate this issue by treating CSI as a visual image, 
but such ``black-box" designs often lack interpretability, producing CSI that is not consistent with multipath propagation principles.
To address these limitations, this paper proposes HS-PINNnet, a \textbf{H}ierarchical \textbf{S}ensing mechanism assisted \textbf{P}hysics-\textbf{I}nformed \textbf{N}eural \textbf{N}etwork for \textbf{C}SI \textbf{F}eedback.
Unlike vision-inspired methods, HS-PINNnet integrates a multipath channel model into the network, reformulating high-dimensional CSI reconstruction as low-dimensional multipath parameter estimation (e.g., amplitude, angle). 
HS-PINNnet features a hierarchical sensing encoder to produce a compact multipath representation, and a heterogeneous decoder for parameter-specific CSI reconstruction, with dedicated branches to estimate different parameters.
Moreover, a PCD module adaptively estimates the number of dominant paths in each CSI sample to enhance generalization across diverse environments. A subchannel-wise shared encoding and parallel decoding strategy is further designed to decompose high-dimensional CSI processing into low-dimensional subchannel tasks, reducing training difficulty and improving scalability of HS-PINNnet for future extremely large-scale multiple-input multiple-output (XL-MIMO) systems.
Simulation results show that HS-PINNnet outperforms the state-of-the-art under different configurations, achieving a 92.8\% reduction in FLOPs and exhibiting two orders of magnitude lower FPGA simulation latency.

\end{abstract}


\section{Introduction}
\label{sec:intro}
In the vision of 6G, ultra-massive multiple-input multiple-output (MIMO) systems serve as a foundational enabler to achieve unprecedented spectral efficiency and system capacity \cite{6gback2}. To support real-time beamforming at base stations (BSs), user equipments (UEs) must continuously report high-dimensional channel state information (CSI) \cite{6gapp}. However, the dimensionality of CSI scales proportionally with the expansion of antennas, resulting in escalating feedback overhead that persistently strains uplink resources \cite{overview}. 
Traditional approaches, including codebook-based quantization \cite{3gppcodebook} and compressive sensing (CS) \cite{cs}, become insufficient for reconstructing high-dimensional channels in such complex scenarios \cite{overview}. This mismatch between growing channel dimensionality and limited capacity poses a critical challenge for efficient CSI acquisition.

\begin{figure}[t]
    \centering
    \includegraphics[width=1\linewidth]{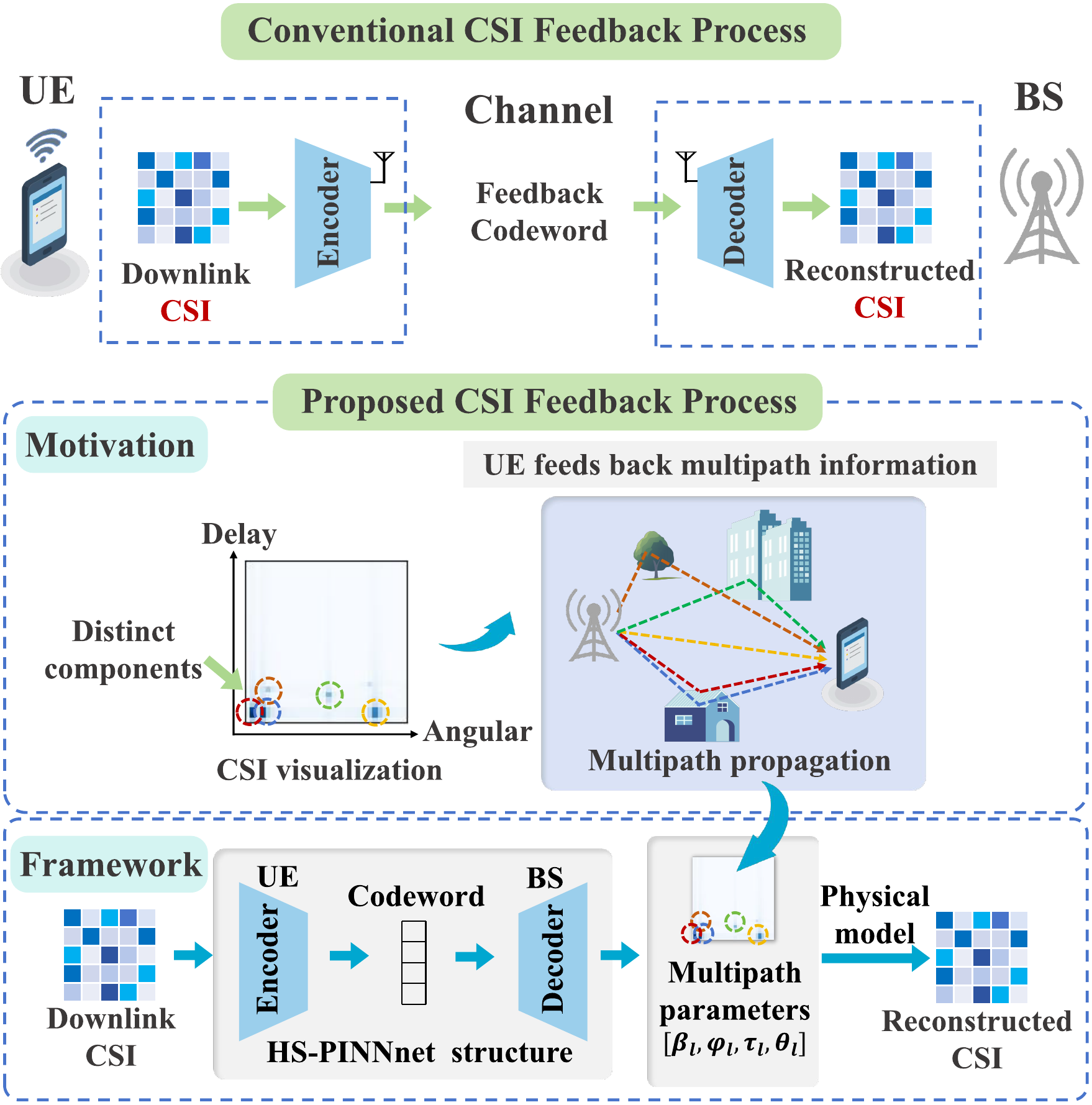}
    \caption{Illustration of conventional CSI feedback and the proposed framework, where a physical model guides the network's learning of CSI physical structure. The encoder compresses multipath features. The decoder recovers multipath parameters. Finally CSI is reconstructed via the model.} 
    \label{fig:motivationframe}
\end{figure}

Recent advancements in deep learning (DL) have significantly improved wireless communication tasks \cite{nnwork1, nnwork2, nnwork3, nnwork4,prompt_AI}. 
Benefiting from the powerful representation capabilities of AI models, explicit CSI feedback has become feasible \cite{cui2026towardsCSIacquisition6G}.
As illustrated in the upper part of Fig. \ref{fig:motivationframe},
DL methods typically treat CSI matrices as two-dimensional (2D) images using autoencoder architectures. CsiNet \cite{csinet} pioneers this paradigm by employing a lightweight convolutional neural network (CNN) encoder and a RefineNet-based decoder.
Subsequent CNN-based methods, such as CsiNet+ \cite{CsiNet+} and CRNet \cite{crnet}, improve reconstruction by enlarging receptive fields or introducing inception-style structures \cite{inception,dual_conv,inception2}. 
Meanwhile, Transformer-based approaches \cite{transformer1,transformer2,TransNet,stnet} leverage attention mechanisms \cite{attention, visiontrans} to model long-range CSI dependencies. Notably, TransNet \cite{TransNet} attains superior accuracy with two-layer Transformer blocks in both encoder and decoder, but incurs substantially high computational complexity.

Despite their effectiveness, these methods largely follow the ``CSI-as-an-image'' paradigm and mainly pursue architectural improvements inspired by computer vision. Such designs often require increasingly complex networks to obtain marginal gains, making it difficult to balance reconstruction performance and computational complexity in real-world systems.

Beyond the ``CSI-as-an-image'' framework, several studies have explored CSI physical properties to improve feedback performance. 
Temporal correlation from slow environmental variation has been exploited in \cite{csinetlstm}.
Frequency correlation has been utilized by modeling subcarrier correlation \cite{frequency1} or leveraging uplink-downlink channel reciprocity to mitigate pilot aliasing and reconstruction errors \cite{bidirectional2,Upsampling}. 
Spatial correlation has been investigated in \cite{pinn_data}, which exploits angular-domain characteristics, and in \cite{cocsinet}, which uses magnitude similarity among nearby UEs to share path information. 
Beyond intrinsic CSI correlations, \cite{MuSAC} exploits the mutualistic correlation between sensing and communication data by embedding sensory information into CSI feedback packets.

Although these works demonstrate the usefulness of physical properties, they mainly treat such properties as statistical priors or correlation patterns, while the intrinsic physical multipath structure of CSI remains underexploited. 
To bridge this gap, \cite{stripeformer} exploits stripe-like CSI patterns, but remains limited to structure modeling at image level. 
The Newtonized Orthogonal Matching Pursuit (NOMP) algorithm \cite{NOMP_algorithm} directly estimates multipath parameters by iteratively selecting and refining atoms from a 2D grid, but incurs high complexity.
\cite{parameterDL} compresses CSI into geometric parameters for direct feedback. While valuable for its focus on limited scattering properties of CSI, it suffers from heavy reliance on precise parameter estimation at UE, insufficient exploration of parameter compressibility and high Transformer-induced complexity.

Motivated by these observations, this work departs from ``CSI-as-an-image'' paradigm and shallow uses of physical priors, by directly incorporating the deterministic multipath propagation model into CSI feedback.
Specifically, CSI is generated by electromagnetic propagation, where signals interact with objects such as buildings and trees and arrive at the receiver through a limited number of paths, each characterized by complex gain, delay, and direction.
This multipath propagation allows CSI to be compactly represented as a superposition of a few dominant path components.

Based on this insight, we introduce HS-PINNnet, a \textbf{H}ierarchical \textbf{S}ensing mechanism assisted \textbf{P}hysics-\textbf{I}nformed \textbf{N}eural \textbf{N}etwork for \textbf{C}SI \textbf{F}eedback, which guides the network to recover the key parameters of propagation paths,
providing a step toward physics-informed wireless representation learning \cite{foundationmodel}.
Compared with direct parameter-feedback schemes \cite{parameterDL}, whose performance is limited by UE-side estimation errors and a fixed number of fed-back dominant paths, HS-PINNnet compresses multipath information into a learned codeword and performs parameter recovery at the BS. Under the same feedback overhead, this codeword can implicitly retain more information beyond a small set of selected paths, enabling HS-PINNnet to outperform practical direct parameter-feedback schemes while reducing UE-side complexity.
The main contributions are summarized as follows: 
\begin{itemize}
    \item{\textbf{Reformulating CSI reconstruction as multipath parameter estimation:} We break away from the ``CSI-as-an-image'' paradigm by exploiting the physical meaning of CSI. Since CSI can be characterized by path parameters such as delays, angles, and amplitudes, the high-dimensional CSI reconstruction problem is reformulated as low-dimensional multipath parameter estimation. A novel method named HS-PINNnet is proposed, which embeds a multipath propagation model into the neural network, providing a step toward physics-informed wireless representation learning.}
    \item{\textbf{Hierarchical sensing convolution for path-aware CSI compression:} For efficient CSI compression in HS-PINNnet, the encoder is designed to identify critical multipath components and produce compact representation. Specifically, a hierarchical sensing convolutional mechanism is developed to replace conventional convolutions, where large kernels perform rough path scanning and small kernels conduct fine-grained path localization, thereby generating high-quality feedback codewords.}
    \item{\textbf{Heterogeneous decoder for parameter estimation with physics-informed optimization:} 
    Under the parameter estimation framework, dedicated decoder branches are designed to handle heterogeneous distributions and estimation difficulties of different parameters, which is supported by theoretical analysis on cumulative reconstruction errors across antennas and subcarriers. The estimated parameters are then used to reconstruct CSI via a physical model, ensuring consistency with multipath propagation principles.
    Meanwhile, a two-stage training strategy is tailored for the proposed structure by first guiding parameter learning and then refining CSI reconstruction.}
    \item{\textbf{Adaptive output path-count configuration and low-complexity large-array processing:} A PCD module is designed to estimate the dominant path-count level at the UE and feed it back to the BS, adjusting the decoder output dimension according to the propagation complexity of each CSI sample. For larger BS arrays, a subchannel-wise shared encoding and parallel decoding strategy decomposes high-dimensional CSI processing into low-dimensional subchannel tasks, reducing training difficulty and demonstrating the scalability of the proposed framework for future extremely large-scale multiple-input multiple-output (XL-MIMO) systems.}

\end{itemize}

The remainder of this paper is organized as follows. Section \ref{sec:system} introduces the system model. Section \ref{sec:method} elaborates the motivation and design of HS-PINNnet. Section \ref{sec:PCD} demonstrates the PCD module for path number estimation. Section \ref{sec: subchannel} introduces the subchannel-wise shared encoding and parallel decoding strategy for larger array processing. Simulation results are provided in Section \ref{sec: results}, and Section \ref{sec:conclusion} concludes the paper.

Notations: In this paper, italic letters denote scalars, while boldface lowercase and uppercase letters represent vectors and matrices, respectively. The Hermitian transpose of a vector or matrix is denoted by $(\cdot)^H$. The regular transpose is indicated by $(\cdot)^T$. $\mathbb{C}^{m\times n}$ ($\mathbb{R}^{m\times n}$) denotes the space of $m \times n$ complex-valued (real-valued) matrices. Euclidean norm of a vector is given by $\|\cdot\|_2$. The magnitude of a scalar is given by $\left| \cdot \right|$. $\mathbb{E}(\cdot)$ is the expectation over a batch of training samples. $\mathbf{A}_{i,j}$ represents the $(i, j)$-th element of matrix $\mathbf{A}$. All other notation is introduced contextually within the text.

\section{System Model}
\label{sec:system}
In this paper, we investigate a typical cellular communication system, the BS is equipped with a uniform linear array comprising  $N_{\mathrm{t}}$ antennas, while the UE employs a single antenna. 
To facilitate spectrum utilization and mitigate inter-symbol interference, orthogonal frequency division multiplexing (OFDM) is adopted with $N_{\mathrm{c}} $ subcarriers.
The received signal at the UE on the $k$-th subcarrier can be expressed as:
\begin{equation}
    y_k = \mathbf{h}_k^H \mathbf{v}_k x_k + n_k,
    \label{eq:yk}
\end{equation}
For the $k$-th subcarrier, $x_k \in \mathbb{C}$ represents the transmitted signal, $\mathbf{v}_k \in \mathbb{C}^{N_{\mathrm{t}} \times 1}$ denotes the precoding vector, $\mathbf{h}_k \in \mathbb{C}^{N_{\mathrm{t}} \times 1}$ is the frequency-domain channel vector, and $n_k \sim \mathcal{N}(0, \sigma_n^2)$ corresponds to the additive Gaussian noise. The CSI matrix $\mathbf{H} \in \mathbb{C}^{N_{\mathrm{c}} \times N_{\mathrm{t}}}$ in frequency-spatial domain is defined as $\mathbf{H} =[\mathbf{h}_1,\dots , \mathbf{h}_{N_\mathrm{c}}]^H$.
Crucially, due to the limited number of scattering paths in massive MIMO environments, $\mathbf{H}$ exhibits significant sparsity when transformed into the angular-delay domain $\tilde{\mathbf{H}}$ via inverse discrete Fourier transformation (IDFT), following \cite{csinet}: 
\begin{equation}
    \tilde{\mathbf{H}} =\mathbf{F}_\mathrm{d} \mathbf{H} \mathbf{F}_\mathrm{a},
    \label{eq:2dft}
\end{equation}
where $\mathbf{F}_\mathrm{d} \in  \mathbb{C}^{N_\mathrm{c} \times N_\mathrm{c}}$ and $\mathbf{F}_\mathrm{a} \in  \mathbb{C}^{N_\mathrm{t} \times N_\mathrm{t}}$ are DFT matrices.
This fundamental physical property is the underlying basis for compression and feedback overhead reduction.

As illustrated in Fig. \ref{fig:motivationframe}, we assume that the UE perfectly acquires the downlink CSI matrix $\mathbf{H}$. In the CSI feedback process, the UE compresses the high-dimensional $\mathbf{H}$ (often processed in its sparse representation $\tilde{\mathbf{H}}$ \cite{csinet}) into a low-dimensional codeword $\mathbf{c}$, which is then transmitted back to the BS. The process can be expressed as:
\begin{equation}
    \mathbf{c}=f_\mathrm{en}(\tilde{\mathbf{H}}),
\end{equation}
where $f_\mathrm{en}$ denotes the encoder at the UE side. 
The BS subsequently reconstructs the CSI matrix $\hat{\mathbf{H}}$ from the received codeword. The process can be expressed as:
\begin{equation}
    \hat{\mathbf{H}}=f_\mathrm{de}(\mathbf{c}),
\end{equation}
where $f_\mathrm{de}$ denotes the decoder at the BS side. 
Thus, the primary objective of CSI feedback is to minimize overhead while ensuring high-fidelity reconstruction of the CSI.

\begin{figure*}[!t]
    \centering
    \includegraphics[width=0.95\linewidth]{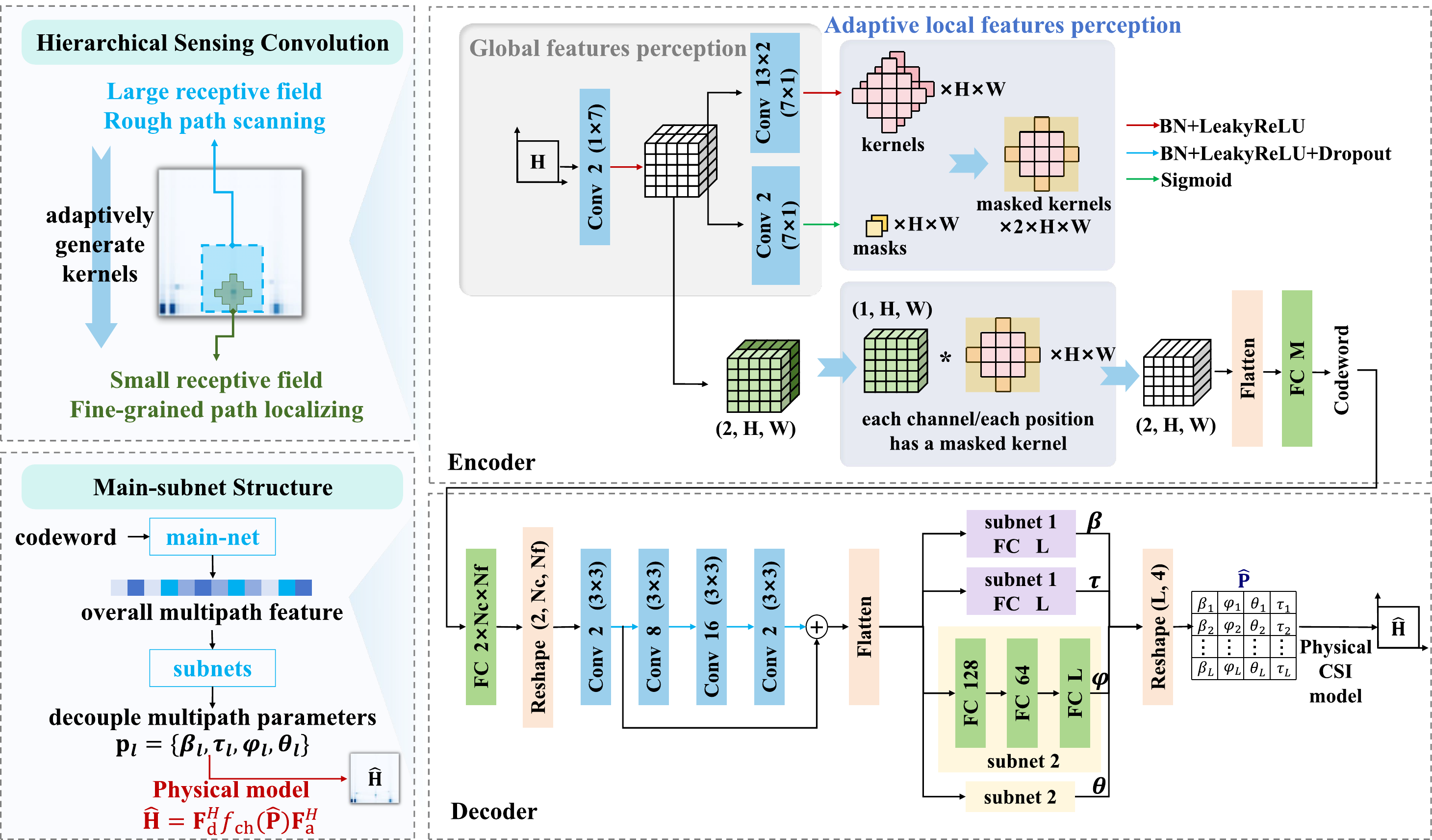}
    \caption{The architecture of HS-PINNnet. The encoder is designed with a hierarchical sensing convolution mechanism, with large kernels leveraging global CSI features and roughly scanning the path region, then enabling adaptive masked small kernels to perceive local details and refining the path localizing. The decoder is designed as a main and subnet structure, recovering the overall multipath features and decoupling each parameter.}
    \label{fig:network}

\end{figure*}
\section{Physics-Informed Neural Network for CSI Feedback}
\label{sec:method}
\subsection{Motivation} 

Current autoencoder-based methods often treat the CSI matrix similarly to how one might treat an image in computer vision tasks, essentially as a grid of pixels, where each element is processed like an independent pixel. These methods typically reconstruct the full-dimensional channel directly using objectives such as mean squared error (MSE), while largely ignoring the physical propagation process that generates CSI. As a result, such black-box compression strategies fail to exploit the inherent physical structure of wireless channels, leading to inefficient and complex models.

However, the downlink CSI matrix is not merely an arbitrary 2D data array, but a deterministic manifestation of the physical propagation environment. It is mainly governed by a limited number of multipath components, each characterized by physical parameters such as angle of departure, delay, and path amplitude. 
Hence, despite its high-dimensional representation, CSI can be compactly described as the superposition of a few dominant propagation paths.

Therefore, as shown in Fig. \ref{fig:motivationframe}, we advocate a physics-informed CSI feedback approach, where ``physics-informed'' indicates that a differentiable channel model is embedded into the neural reconstruction and optimization pipeline. Instead of directly compressing CSI as an image, the proposed framework estimates the lower-dimensional physical causes of CSI, namely the dominant multipath parameters, and reconstructs CSI through the channel model. 
This design improves physical consistency in terms of model-guided output generation, aligning CSI reconstruction with wireless propagation physics. Meanwhile, such a framework reduces feedback overhead by transforming the task from a high-dimensional image-level recovery to a compact parameter-level estimation, providing a promising step toward physics-informed wireless representation learning with improved interpretability.

\subsection{Framework Overview}
Since the physical meaning of CSI is fundamentally embedded within its distinct multipath components, an effective representation for these components is essential.
Thus, $\mathbf{h}_k$ of CSI can be approximated by a stochastic geometric physical channel model, following: 
\begin{equation} 
    \mathbf{h}_k = \sum_{l=1}^{L} \alpha_l e^{-j 2\pi (k-1) \Delta f  \tau_l} \mathbf{a}(\theta_l),
    \label{eq:hk}
\end{equation}  
where $\mathbf{a}(\theta_l)=\left[1, e^{j \frac{2 \pi d \sin(\theta_l)}{\lambda }}, \ldots, e^{j (N_\mathrm{t}-1) \frac{2 \pi d \sin(\theta_l)}{\lambda } }\right]^{T}$ is the array steering vector across $ N_\mathrm{t}$ antennas. $L$ is the number of propagation paths and $\Delta f$ is the subcarrier spacing. $d$ and $\lambda$ denote the antenna spacing and the signal wavelength, respectively. $\theta_l \in [0,2\pi)$ is the angle of departure of the $l$th path, and $\tau_l \in [0,\tau_\mathrm{max}]$ denotes the time delay. The complex gain $\alpha_l=\beta_l e^{j\phi_l}$ is composed of path amplitude $\beta_l \in [0,\beta_\mathrm{max}]$ and phase $\phi_l \in [0,2\pi)$. 
The parameters $\tau_\mathrm{max}$ and $\beta_\mathrm{max}$ represent the maximum delay and maximum path amplitude of all paths. Then, the downlink CSI matrix $\mathbf{H}$ can be expressed using the physical multipath model:
\begin{equation}
    \mathbf{H} =[\mathbf{h}_1,\dots , \mathbf{h}_{N_\mathrm{c}}]^H=\sum_{l=1}^{L} \beta_l e^{j\phi_l}  \mathbf{b}(\tau_l)\mathbf{a}^H(\theta_l),
    \label{eq:H}
\end{equation}
where $\mathbf{b}(\tau_l)=\left[1, e^{-j 2 \pi \Delta f \tau_l}, \ldots, e^{-j (N_\mathrm{c}-1) 2 \pi \Delta f \tau_l }\right]^{T}$ is the frequency response vector across $N_\mathrm{c}$ subcarriers.

Equation (\ref{eq:H}) provides an effective representation of the multipath information, where each propagation path can be characterized by the multipath parameter set $\mathbf{P}=[\mathbf{p}_1,\ldots,\mathbf{p}_L]^T \in \mathbb{R}^{L \times 4}$. The parameters $\mathbf{p}_l$ of the $l$th path are defined as
\begin{equation}
    \mathbf{p}_l=\left \{ \beta_l,\phi_l,\theta_l,\tau_l\right \}. 
\end{equation}
Based on this analysis, it is sufficient to feed back the multipath features and reconstruct the parameter set $\mathbf{P}$ at the BS to reconstruct the downlink CSI.

Therefore, we propose a physics-informed CSI feedback framework based on the signal propagation multipath model. As shown in Fig. \ref{fig:motivationframe}, the encoder at the UE and the decoder at the BS are implemented using deep neural networks (DNNs). The encoder compresses the multipath parameter features of the CSI into a codeword $\mathbf{c}$, and the decoder recovers the exact multipath parameters $\mathbf{P}$ from the received codeword $\mathbf{c}$. 

We first transform the CSI matrix $\mathbf{H}$ in (\ref{eq:H}) from the frequency-spatial domain to the angular-delay domain $\tilde{\mathbf{H}}$ via IDFT in (\ref{eq:2dft}), whose sparse representation facilitates the network's learning of the fundamental multipath characteristics inherent in the CSI.
The complex-valued CSI matrix $\tilde{\mathbf{H}}$ is then rearranged into a real-valued tensor $\mathbf{H}_\mathrm{in} \in \mathbb{R}^{2 \times N_\mathrm{t} \times N_\mathrm{c}}$ to serve as the input to the encoder. The process of encoding $\mathbf{H}_\mathrm{in}$ into a codeword $\mathbf{c} \in \mathbb{R}^{M \times 1}$ is expressed as
\begin{equation}
    \mathbf{c}=f_\mathrm{en}(\mathbf{H}_\mathrm{in},\boldsymbol{\Theta}_\mathrm{en}),
\end{equation}
where $f_\mathrm{en}$ denotes the encoder and $\boldsymbol{\Theta}_\mathrm{en}$ its parameters. As shown in Fig. \ref{fig:motivationframe}, the decoder at the BS recovers the multipath parameters $\hat{\mathbf{P}} \in \mathbb{R}^{L \times 4}$ from the codeword $\mathbf{c}$: 
\begin{equation}
    \hat{\mathbf{P}}=f_\mathrm{de}(\mathbf{c},\boldsymbol{\Theta}_\mathrm{de}),
\end{equation}
where $f_\mathrm{de}$ and $\boldsymbol{\Theta}_\mathrm{de}$ represent the decoder and its network parameters, respectively. The output $\hat{\mathbf{P}} = [\hat{\mathbf{p}}_1,\ldots,\hat{\mathbf{p}}_L]^T$ contains all recovered parameters across the $L$ paths. Finally, the reconstructed CSI $\mathbf{\hat{H}}$ can be obtained by:
\begin{equation}
    \mathbf{\hat{H}}=\mathbf{F}_\mathrm{d}^H f_\mathrm{ch}(\hat{\mathbf{P}})\mathbf{F}_\mathrm{a}^H,
    \label{eq:reconstrcut_CSI}
\end{equation}
where $f_\mathrm{ch}$ denotes the channel reconstruction process based on (\ref{eq:H}).
Guided by this framework, we propose HS-PINNnet, which recovers CSI according to the estimated multipath parameters. 
Compared with direct parameter-feedback schemes \cite{parameterDL}, whose performance is affected by UE-side parameter estimation errors and constrained by the fixed number of fed-back dominant paths even under perfect estimation, HS-PINNnet compresses multipath information into a learned codeword and offloads parameter recovery to the BS. Thus, under the same feedback overhead, the codeword can implicitly preserve information of more propagation components beyond a small set of selected dominant paths. This enables HS-PINNnet to outperform practical direct parameter-feedback schemes with imperfect UE-side estimation. Details are described in the following sections.

\subsection{HS-PINNnet Architecture}
HS-PINNnet is designed with a hierarchical sensing convolutional encoder and a hetero-architecture decoder with a main and subnet structure, as illustrated in Fig. \ref{fig:network}. In the following sections, we introduce the encoder and decoder, respectively.

\subsubsection{Hierarchical Sensing Convolutional Encoder}
The encoder aims to detect and compress multipath features into a low-dimensional codeword for efficient decoding. Since dominant multipath parameters are mainly derived from localized regions in the angular-delay CSI, the encoder should emphasize path-dominant areas while suppressing irrelevant or noisy components. 
However, conventional convolutions share kernels across the entire input and treat all spatial locations with uniform sensitivity, which is inefficient because only limited angular-delay regions contain meaningful multipath information as shown in Fig. \ref{fig:motivationframe}. This motivates region-specific feature extraction to enhance path-related components over background regions \cite{Deformable_conv}.

To this end, we propose a hierarchical sensing convolutional encoder that combines the complementary strengths of large and small kernels \cite{large_kernel,Combine_kernel,LSNet}. 
Specifically, large kernels, due to their broader receptive fields, are well suited for scanning dominant multipath components, while small kernels, with localized receptive fields, are more appropriate for capturing fine-grained multipath details.
Accordingly, the hierarchical sensing convolutional encoder first applies large kernels to extract global CSI features and coarsely identify path regions, then generates position-specific small kernels with location-dependent values on the feature map, enabling fine-grained refinement of multipath localization.

The above process is illustrated in Fig. \ref{fig:network}.
In this architecture, we adopt asymmetric large-kernel convolution \cite{cnnsurvey} to independently capture features in the angular and delay domains. A convolutional layer with a $1 \times 7$ kernel is initially applied to guide the network in learning global features and generating intermediate feature maps.
\begin{figure}[!t]
    \centering
    \includegraphics[width=1\linewidth]{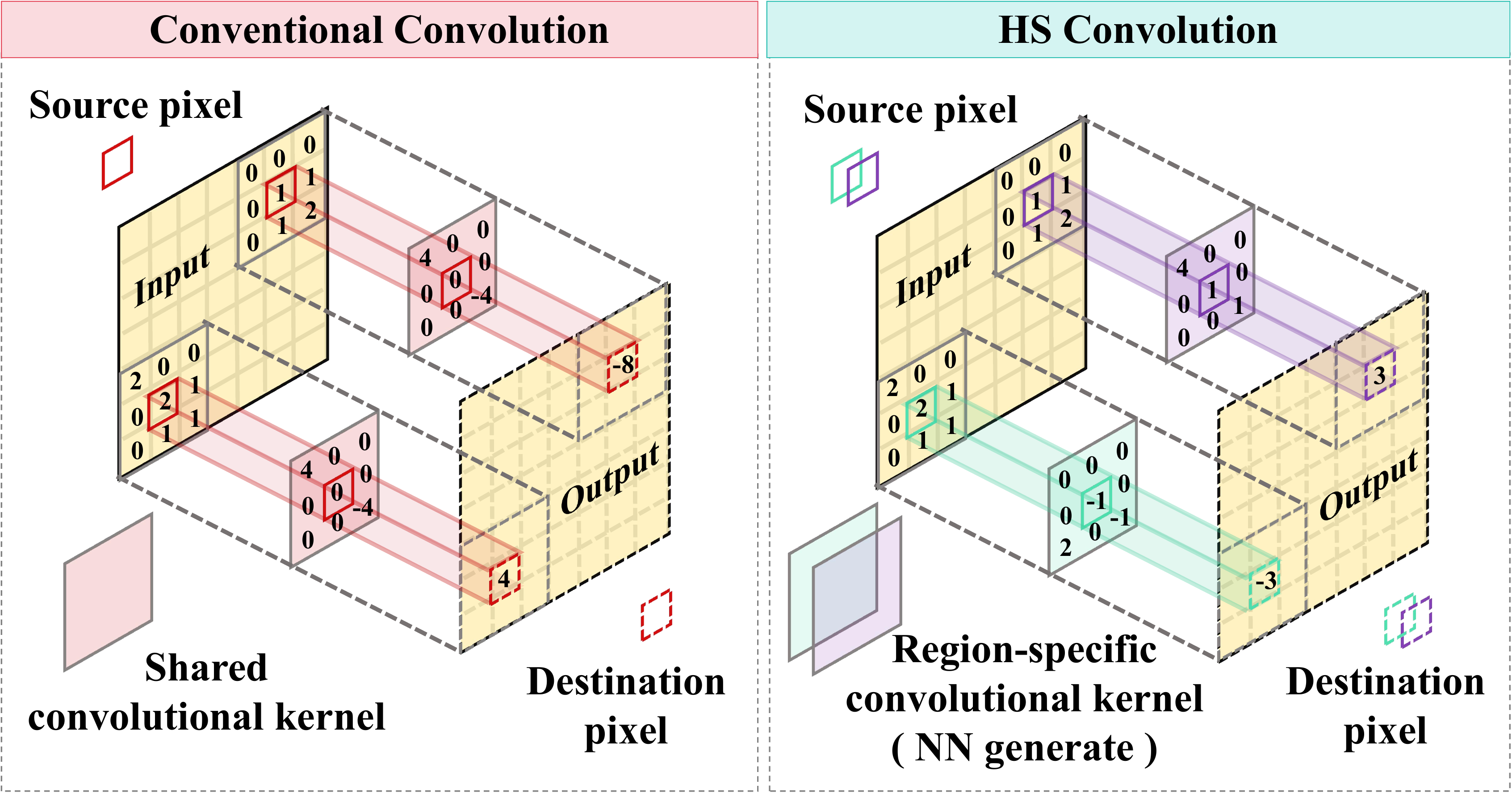}
    \caption{The difference between conventional convolution and HS convolution. For clarity in representation, the region-specific convolutional kernels is shown as a $3 \times 3$ format.}
    \label{fig:regionConv}
\end{figure}
The encoder is then split into two branches, each using a $7 \times 1$ kernel. One branch generates position-specific diamond-shaped kernels $\mathbf{K} \in \mathbb{R}^{2\times H\times W \times k_\text{s}}$, where $k_\text{s}$ is the kernel size. For ease of illustration, we use $\mathbf{K}_{c,h,w,i,j}$ to denote each element of the generated kernel, where $c$ is the feature channel, $h$ and $w$ indicate the spatial position on the intermediate feature map, and $i,j \in \left\{ 0,1,2,3,4 \right\}$ are the local kernel indices centered at $(2,2)$. To match the stripe-like patterns caused by the DFT windowing effect, only positions satisfying $|i-2|+|j-2|\leq2$ are assigned valid values, while other positions are set to zero.

Due to the different levels of angular and delay spread observed across paths, the other branch is used to generate binary masks $\mathbf{m} \in \left\{0,1\right\}^{2\times H\times W}$ that control the activation strength across surrounding kernel regions. These masks allow the network to adapt to varying spatial spreads by modulating the extent of feature extraction. We combine the masks with the diamond-shaped kernels to generate masked kernels $\tilde{\mathbf{K}} \in \mathbb{R}^{2 \times H\times W \times k_\text{s}}$, the process can be expressed as:
\begin{equation}
   \tilde{\mathbf{K}}_{c,h,w,i,j}=
   \begin{cases}
   \mathbf{m}_{c,h,w}\mathbf{K}_{c,h,w,i,j}& \text{if }|i-2|=2\text{ or }|j-2|=2\\
   \mathbf{K}_{c,h,w,i,j} & \text{otherwise}
   \end{cases}
\end{equation}

Subsequently, for each position in the $2 \times H \times W$ grid of the intermediate feature map, a unique masked kernel is applied to obtain the region-specific convolutional output, as shown in Fig. \ref{fig:regionConv}. 
\begin{figure*}
    \centering
    \includegraphics[width=0.95\linewidth]{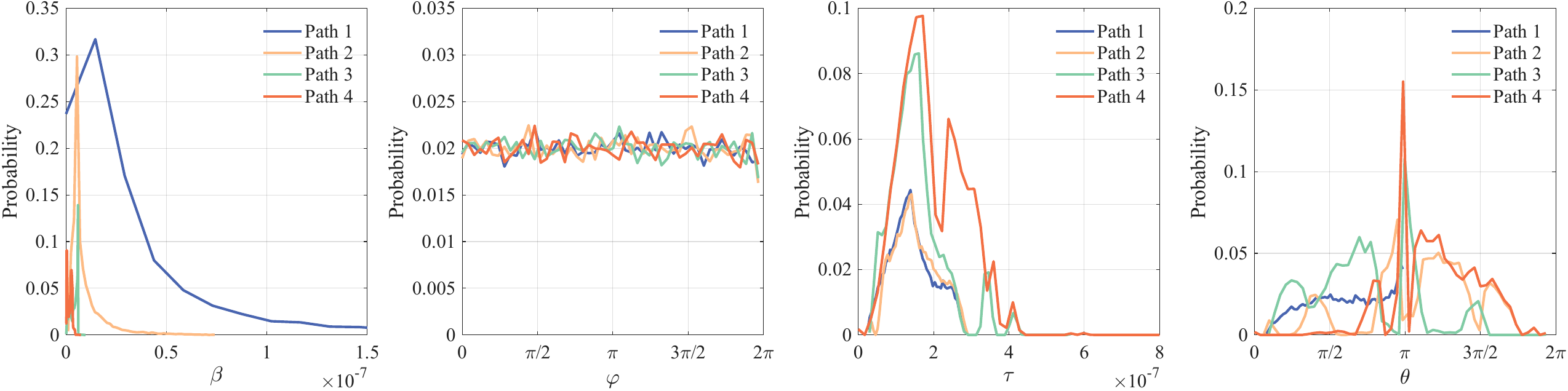}
    \caption{Parameter distribution of the simulations' scenario. $\beta$ and $\tau$ are relatively concentrated with an obvious peak and comparatively easier to estimate. The phase $\phi$ tends toward a uniform random distribution and the antenna angle $\theta$ exhibits a different distribution for each path. These two parameters are inherently stochastic and more challenging to predict. }
    \label{fig:parameterdis}
\end{figure*}
Finally, a fully connected (FC) layer transforms the output into a low-dimensional codeword $\mathbf{c}$ of length $M$. This codeword is then transmitted from the UE to the BS.

\subsubsection{Main and Subnet Decoder}
The hetero-architecture decoder consists of a shared main network and a branched subnet for estimating the multipath parameter matrix $\hat{\mathbf{P}}$, as shown in Fig. \ref{fig:network}. 
The main network serves as a feature extractor, tasked with extracting abstract features from the input codeword $\mathbf{c}$ through stacked convolutional layers.
Batch normalization \cite{BN_layer} is inserted to stabilize feature distributions and accelerate convergence. A residual learning structure \cite{He_2016_CVPR} is further adopted to alleviate degradation in deep networks.

The extracted feature is then fed into the subnet to estimate and decouple individual multipath parameters. Since $\beta_l$, $\phi_l$, $\theta_l$, and $\tau_l$ exhibit heterogeneous distributions and estimation difficulties, the subnet employs a multi-branch architecture, where each branch is tailored to one parameter. 
This branching strategy is informed by a detailed analysis of the parameters’ statistical distribution and  network computational requirements, optimizing both accuracy and efficiency in parameter reconstruction. Fig. \ref{fig:parameterdis} illustrates the distribution of multipath parameters and the following physical and empirical considerations guide the architectural design:

\begin{itemize}
    \item{$\beta_l$ determines the strength of the corresponding multipath component and appears as a prominent intensity feature in the multipath response. As shown in Fig. \ref{fig:parameterdis}, $\beta_l$ values are relatively concentrated with clear peaks, making them suitable for estimation using a lightweight branch.}
    \item{$\phi_l$ is influenced by both the real and imaginary CSI components and exhibits circular periodicity and discontinuities. Meanwhile, its nearly uniform distribution over $[0,2\pi)$ indicates strong stochasticity. Thus a deeper branch is adopted for $\phi_l$ estimation.}
    \item{$\theta_l$ determines the spatial direction of each path and the position of energy concentration along the angular axis. Due to environmental scattering, $\theta_l$ varies significantly across samples and paths, covering the full angular range, and thus requires a deeper branch for accurate estimation.}
    \item{$\tau_l$ controls the distribution along the delay axis and is typically bounded by the maximum propagation delay. As shown in Fig. \ref{fig:parameterdis}, $\tau_l$ has a concentrated distribution with clear peaks, making it easier to estimate.}
\end{itemize}

Thus to reflect these differences in estimation difficulty, different branch depths are assigned according to parameter estimation difficulty: lightweight single-layer FC branches are used for $\beta_l$ and $\tau_l$, whereas deeper three-layer FC branches are used for $\phi_l$ and $\theta_l$. The output dimension $L$ of each branch corresponds to the estimated number of dominant propagation paths, which can be determined based on PCD module introduced in Section~\ref{sec:PCD}. 
Finally, the branch outputs are concatenated into a structured parameter matrix $\hat{\mathbf{P}} \in \mathbb{R}^{L \times 4}$, representing the recovered multipath parameters. These parameters are then used to calculate the downlink CSI through the channel model in (\ref{eq:H}).

\begin{figure}
    \centering
    \includegraphics[width=0.9\linewidth]{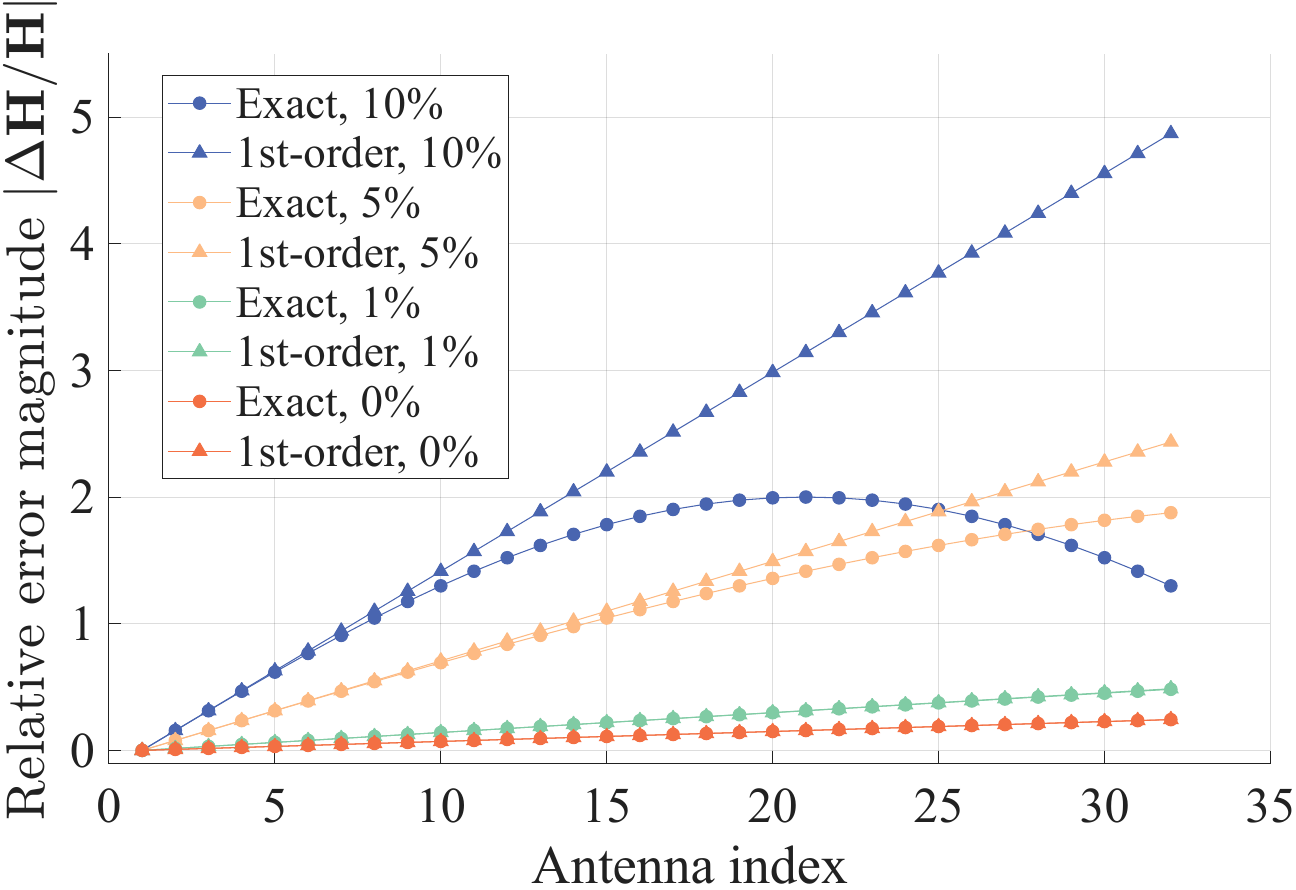}
    \caption{Accumulated error of $\mu$ mismatch ($\mu=\pi/2,N_\mathrm{t}=32$)}
    \label{fig:error1}
\end{figure}
\begin{figure}
    \centering
    \includegraphics[width=0.9\linewidth]{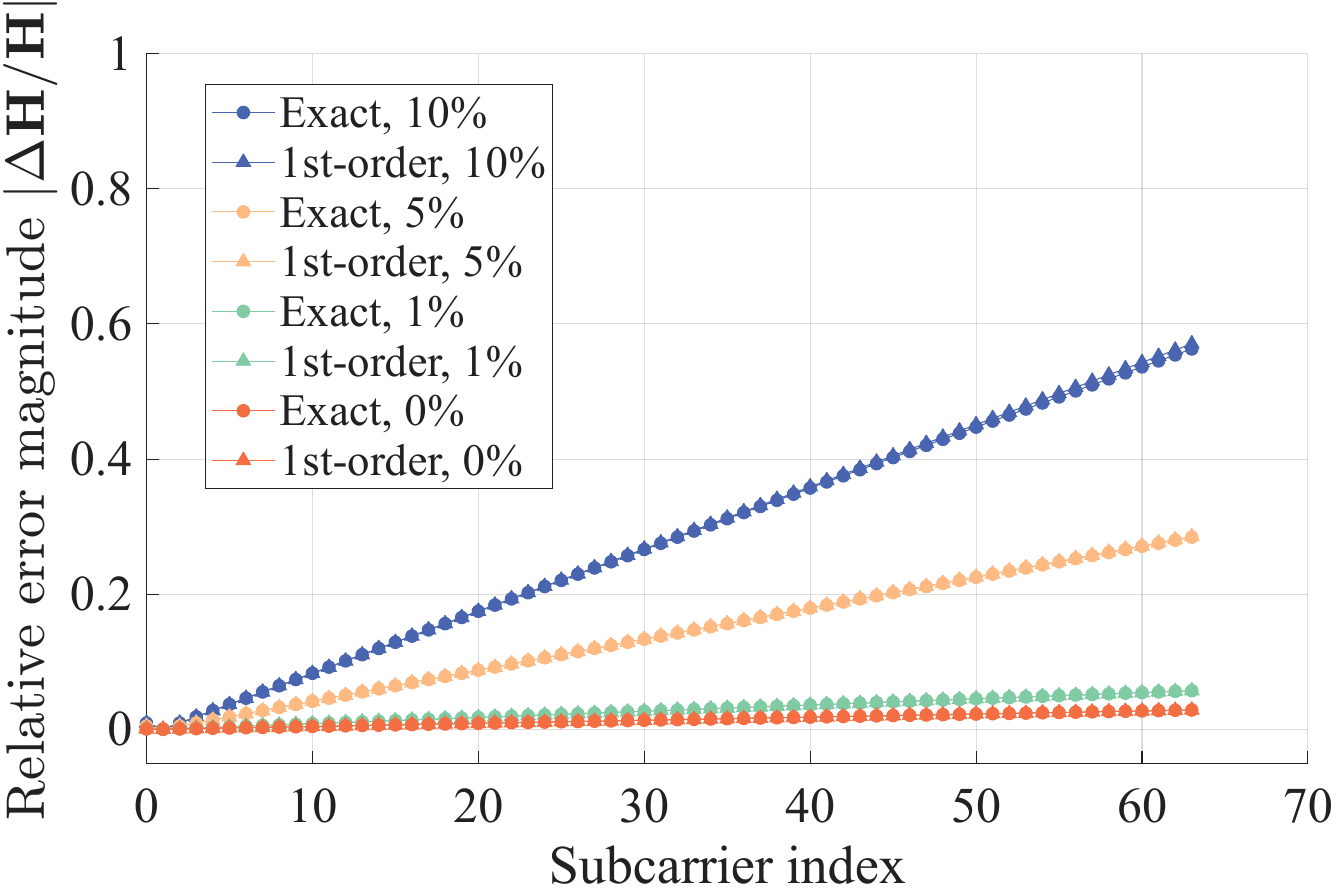}
    \caption{Accumulated error of $\tau$ mismatch ($\tau=1.5 \times10^{-7},N_\mathrm{c}=64$)}
    \label{fig:error2}
\end{figure}
\subsubsection{Theoretical Analysis of Multipath Parameter Estimation Errors}

\label{sec:error analysis}

To quantitatively characterize the impact of multipath parameter estimation errors on the reconstructed CSI, we analyze the perturbation of the channel response caused by mismatches in the path amplitude, phase, delay, and angle parameters.
For a wideband channel with $L$ propagation paths, the channel vector at the $k$-th subcarrier can be written as (\ref{eq:hk}). Accordingly, the contribution of the $l$-th path to the channel coefficient at the $n$-th antenna and the $k$-th subcarrier can be expressed as
\begin{equation}
h_k(n,l)
=\beta_l e^{j\phi_l} e^{-j2\pi (k-1)\Delta f \tau_l} e^{j (n-1) \mu_l},
\end{equation}
where $\mu_l = \frac{2\pi d}{\lambda}\sin\theta_l$ for simplification.
Suppose that the estimated parameters are perturbed as
\begin{equation}
\begin{aligned}
\hat{\beta}_l  &= \beta_l + \Delta\beta_l, &
\hat{\phi}_l   &= \phi_l + \Delta\phi_l, \\
\hat{\tau}_l   &= \tau_l + \Delta\tau_l, &
\hat{\theta}_l &= \theta_l + \Delta\theta_l.
\end{aligned}
\end{equation}
and thus $\hat{\mu}_l=\frac{2\pi d}{\lambda}\sin(\theta_l+\Delta\theta_l)$.
Using the first-order Taylor expansion, the frequency-spatial perturbation can be approximated as
\begin{equation}
\Delta \mu_l=\hat{\mu}_l-\mu_l\approx\frac{2\pi d}{\lambda}\cos\theta_l \, \Delta\theta_l.
\end{equation}
Then, the reconstructed path component becomes    
\begin{equation}
\begin{aligned}
\hat{h}_k(n,l)
&= (\beta_l+\Delta\beta_l)
e^{j(\phi_l+\Delta\phi_l)}\\
&~~\times e^{-j2\pi (k-1)\Delta f (\tau_l+\Delta\tau_l)}e^{j (n-1) (\mu_l+\Delta\mu_l)}.
\end{aligned}
\end{equation}
Applying first-order approximations
\begin{equation}
e^{j\Delta\phi_l}\approx 1+j\Delta\phi_l,
\end{equation}
\begin{equation}
e^{-j2\pi (k-1)\Delta f \Delta\tau_l}
\approx
1-j2\pi (k-1)\Delta f \Delta\tau_l,\end{equation}
\begin{equation}
e^{j(n-1)\Delta\mu_l}\approx 1+j(n-1)\Delta\mu_l,
\end{equation}
we obtain the first-order perturbation of the $l$-th path is
\begin{align}
\Delta h_l(n,k)
&=
\hat{h}_l(n,k)-h_l(n,k) \notag\\
&\approx h_l(n,k)\bigg(
\frac{\Delta\beta_l}{\beta_l}
+j\Delta\phi_l
-j2\pi(k-1)\Delta f\,\Delta\tau_l \notag\\
&\hspace{2.0cm}
+j(n-1)\Delta\mu_l
\bigg).
\label{eq:first_order_error}
\end{align}
Equation~\eqref{eq:first_order_error} reveals that the errors caused by amplitude mismatch $\Delta\beta_l$ and phase mismatch $\Delta\phi_l$ do not accumulate with antenna index $n$ or subcarrier index $k$. In contrast, delay mismatch $\Delta\tau_l$ is multiplied by $(k-1)$, and angle mismatch $\Delta\theta_l$ ($\Delta\mu_l$) is multiplied by $(n-1)$.
To quantify this effect, Fig. \ref{fig:error1} and Fig. \ref{fig:error2} show the element-wise channel error under different relative perturbations of $\tau_l$ and $\mu_l$, with $\tau_l=1.5\times10^{-7}$ s and $\mu_l=\pi/2$. Results show that angular mismatch causes much faster error growth than delay mismatch. For example, with $32$ antennas, a $1\%$ angular error induces nearly $50\%$ distortion, whereas with $64$ subcarriers, a $1\%$ delay error leads to less than $10\%$ deviation.
These observations suggest that angle parameters require higher estimation accuracy than delay parameters, supporting the use of a deeper subnet branch for angle estimation. It also explains why even small parameter errors may cause substantial CSI reconstruction distortion, motivating a training objective that suppresses such amplified errors, as introduced in Section~\ref{sec: strategy}.

\begin{figure*}
    \centering
    \includegraphics[width=1\linewidth]{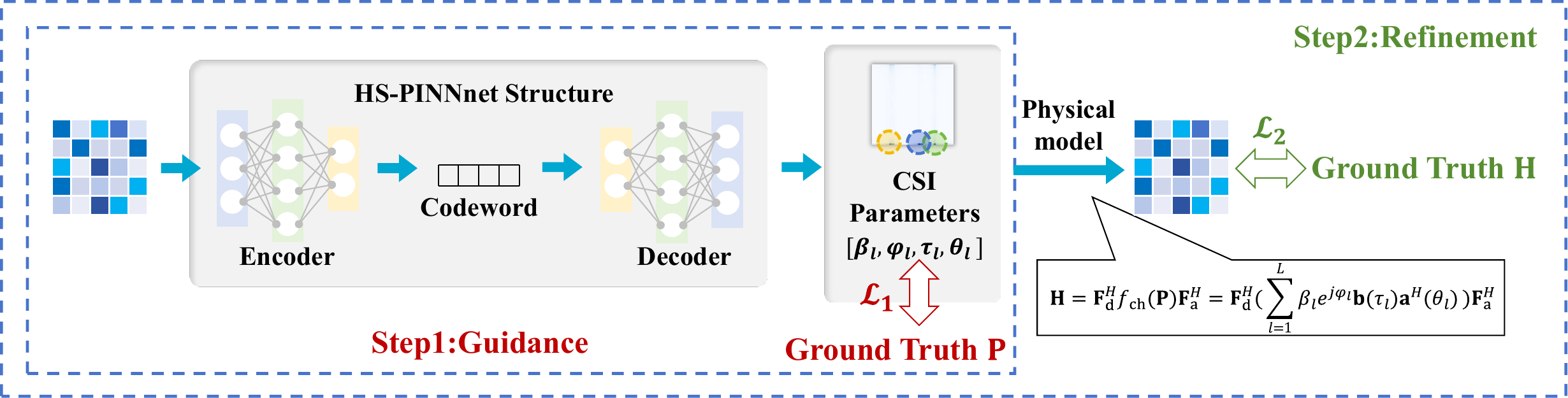}
    \caption{The process of guidance and refinement training strategy. STEP1 guides HS-PINNnet to learn the map between CSI and multipath parameters, better accelerating the convergency of CSI reconstruction. STEP2 fine-tunes the pre-trained model and improves the accuracy of CSI reconstruction.}
    \label{fig:strategy}
\end{figure*}

\subsection{Guidance and Refinement Training Strategy}
\label{sec: strategy}
Conventional DL-based methods typically minimize MSE between the reconstructed and original CSI. In contrast, HS-PINNnet leverages channel physics by taking the CSI matrix as input and the multipath parameters as output. Hence, minimizing parameter error is critical, as it guides the neural network to better capture the intrinsic physics of CSI. However, parameter estimation accuracy alone does not guarantee satisfactory CSI reconstruction performance. As analyzed in section \ref{sec:error analysis}, even small parameter mismatches may accumulate over antenna and subcarrier indices, leading to non-negligible reconstruction distortion, which needs to be further refined by minimizing the reconstruction error of CSI. To address both objectives, we propose a guidance and refinement training strategy, as illustrated in Fig. \ref{fig:strategy}.

\subsubsection{STEP 1: Guidance}
In the first training step, HS-PINNnet is guided to learn the mapping from the input CSI matrix $\mathbf{H}_\mathrm{in}$ to the output parameter matrix $\hat{\mathbf{P}}$ based on the physical multipath model. The goal is to ensure that $\hat{\mathbf{P}}$ closely approximates the ground-truth parameter matrix $\mathbf{P}$. The loss function is defined as the MSE between the predicted and ground-truth parameter matrices: 
\begin{equation}
    \mathcal{L}_{1}(\boldsymbol{\Theta}_\mathrm{en},\boldsymbol{\Theta}_\mathrm{de})=\frac{1}{N_\mathrm{B}} \sum_{i=1}^{N_\mathrm{B}} \|\hat{\mathbf{P}}[i]-\mathbf{P}[i] \|_{2}^{2},
    \label{eq:L1}
\end{equation}
where $N_\mathrm{B}$ is the batch size, and $\boldsymbol{\Theta}_\mathrm{en}$ and $\boldsymbol{\Theta}_\mathrm{de}$ are the parameters of the encoder and decoder. By prioritizing the precise estimation of multipath parameters in this step, HS-PINNnet establishes a foundation for subsequent CSI reconstruction.

\subsubsection{STEP 2: Refinement}
The ultimate goal is to accurately reconstruct the downlink CSI $\mathbf{\hat{H}}$, since optimizing only the multipath parameters does not necessarily lead to accurate reconstructed channels as discussed above. Therefore, in the second training step, we fine-tune the network by optimizing $\hat{\mathbf{P}}$ to better align with the CSI reconstruction model. The reconstructed CSI is calculated in (\ref{eq:reconstrcut_CSI}).
The loss function for this step is defined as the normalized mean squared error (NMSE) between $\mathbf{\hat{H}}$ and the ground-truth CSI $\mathbf{H}$, which naturally incorporates the channel model into the back-propagated optimization as shown in Fig.~\ref{fig:strategy}:
\begin{equation}
    \mathcal{L}_{2}(\boldsymbol{\Theta}_\mathrm{en},\boldsymbol{\Theta}_\mathrm{de})=\frac{1}{N_\mathrm{B}} \sum_{i=1}^{N_\mathrm{B}} \frac{ \|\hat{\mathbf{H}}[i]-\mathbf{H}[i] \|_{2}^{2}}{\left\|\mathbf{H}[i]\right\|_{2}^{2}}.
    \label{eq:L2}
\end{equation} 
We update $\boldsymbol{\Theta}_\mathrm{en}$ and $\boldsymbol{\Theta}_\mathrm{de}$ by applying this two-step training strategy, to jointly learn both an accurate parameter estimator and an effective CSI reconstructor.

\begin{figure}[t]
    \centering
    \includegraphics[width=1\linewidth]{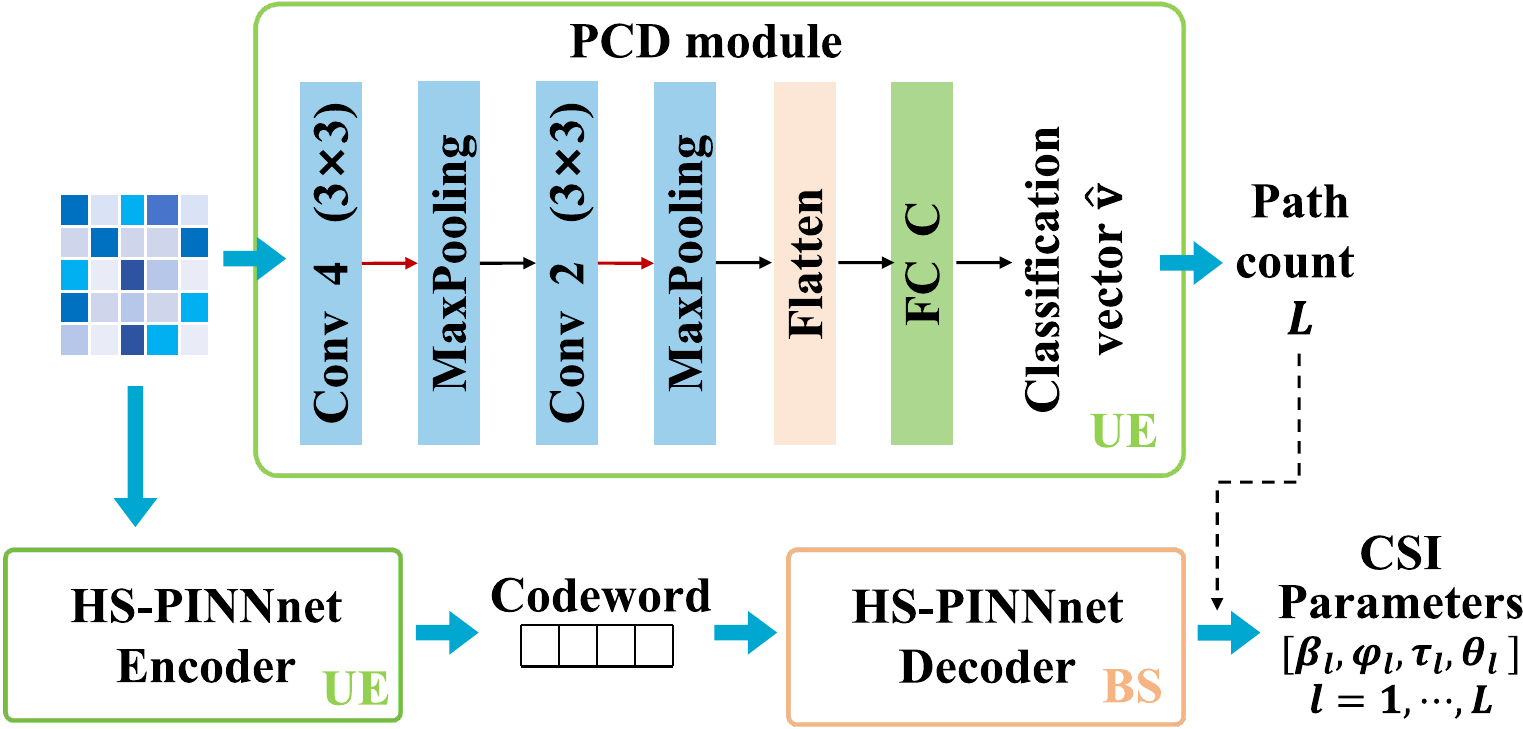}
    \caption{The PCD module estimates the number of dominant paths to be recovered and then feeds it back to BS.}
    \label{fig:PCD}
\end{figure}

\section{Path Count Detection Module for Adaptive Multipath Recovery}
\label{sec:PCD}
Most existing CSI feedback methods adopt a fixed feedback overhead for all channel samples. In such a design, the same latent dimension is used regardless of the actual propagation complexity of the input CSI. However, wireless channels exhibit highly diverse multipath characteristics across users and environments, thus the amount of information contained in different CSI samples is inherently non-uniform. As a result, a fixed-overhead feedback strategy is inefficient.

Motivated by this insight, we propose an adaptive multipath recovery module, where the recovery process is adjusted according to the multipath richness of the input CSI sample. To this end, a lightweight Path Count Detection (PCD) module is deployed at UE side to estimate the number of dominant paths that need to be recovered from the input CSI. The detected path-count class is then fed back to the BS to determine the output dimension $L$ of the subnet. As a result, the proposed CSI feedback framework can adapt to the propagation complexity of each sample, instead of relying on a fixed path-count assumption for all channels. The process is shown in Fig. \ref{fig:PCD}.

\begin{figure*}[t]
    \centering    \includegraphics[width=0.65\linewidth]{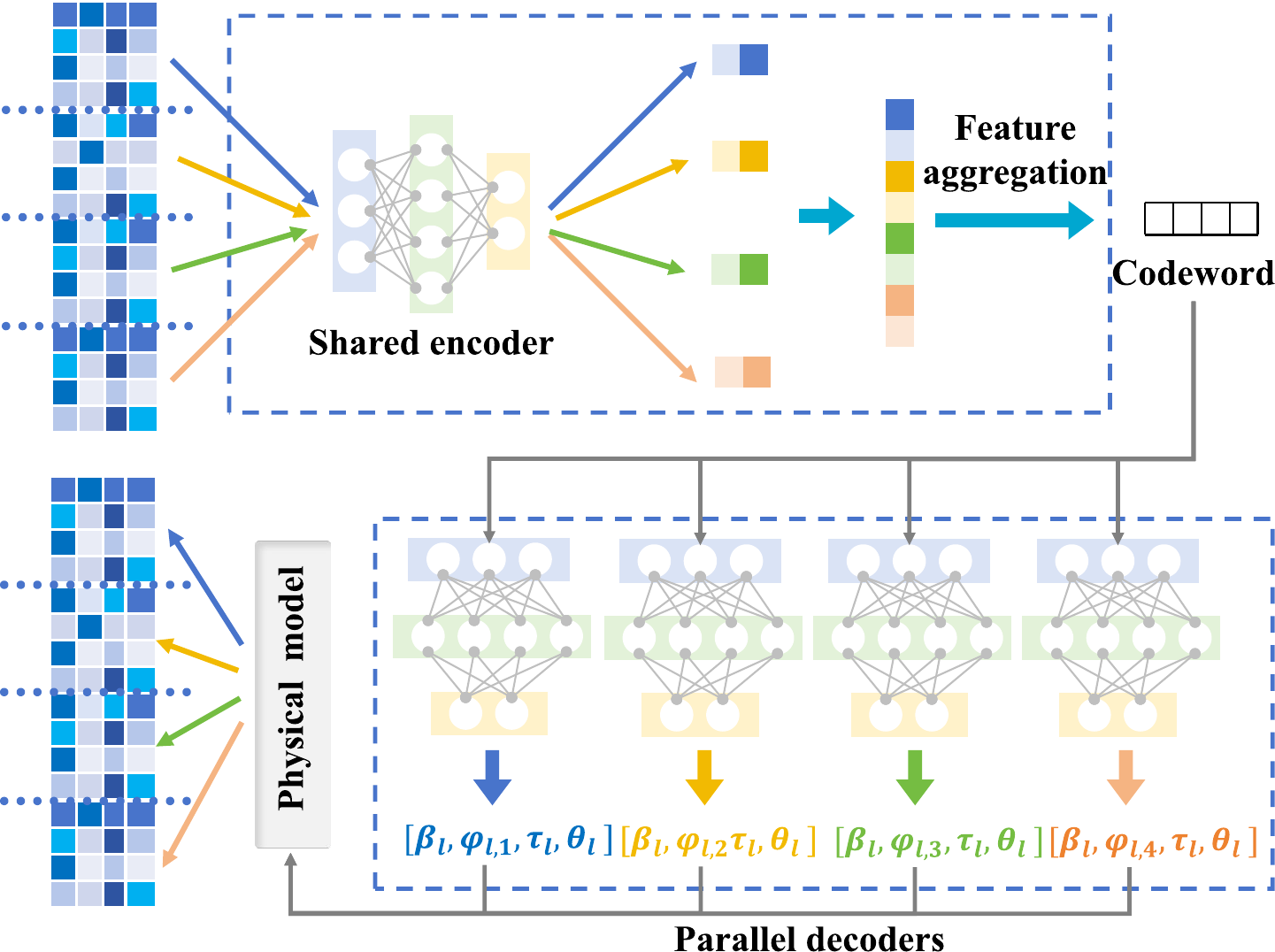}
    \caption{The illustration of subchannel-wise shared encoding and parallel decoding strategy for large BS arrays.}
    \label{fig:subchannel}
\end{figure*}

\subsection{Network Architecture}

The PCD module is designed as a lightweight classification network, which takes $\mathbf{H}_{in}$ as input and outputs a path-count classification vector
\begin{equation}
\hat{\mathbf{v}}=\left[\hat{v}_1,\hat{v}_2,\ldots,\hat{v}_C\right]^{T},
\end{equation}
where $C$ is the number of path-count classes, and $\hat{v}_c$ denotes the predicted probability that the input CSI belongs to the $c$-th class. The detected class is then mapped to the path number $L$ used by the parameter recovery network at BS through $f_\text{c2L}$. The process can be explained by:
\begin{equation}
L=f_\text{c2L}(\operatorname{argmax}(\hat{\mathbf{v}}))=f_\text{c2L}(\operatorname{argmax}(f_{\text{PCD}}(\mathbf{H}_\mathrm{in})))
\end{equation}

As shown in Fig.\ref{fig:PCD}, the network is intentionally kept simple. The input CSI is first processed by a convolutional layer with $4$ filters, followed by batch normalization and LeakyReLU activation. A max-pooling layer is then applied to reduce the spatial resolution and suppress feature redundancy. Next, a convolutional layer with $2$ filters, followed by batch normalization, LeakyReLU, and max-pooling, is employed to further compress the representation. The resulting feature maps are flattened into a one-dimensional vector and passed to a fully connected layer, which outputs the final classification logits.

\subsection{Training Process}

The PCD module is trained as a multi-class classifier. To improve the separability among different classes, we do not assign one class to every exact path number. Instead, the labels are grouped with a step of $2$ paths. Specifically, samples with $L_\text{true}=1$ and $L_\text{true}=2$ are assigned to the class corresponding to detecting $L=2$ paths, samples with $L_\text{true}=3$ and $L_\text{true}=4$ are assigned to the class corresponding to detecting $L=4$ paths, and so on. This two-path grouping is adopted for the reason that if every possible path number were treated as an individual class, the classification boundaries would become overly dense, making the training process more difficult and reducing inter-class discriminability. Therefore, the classification space is instead coarsened into several representative path-count levels.

Let $\mathbf{q}=\left[q_1,q_2,\ldots,q_C\right]^T$ denote the one-hot target label corresponding to the grouped path-count class, where
\begin{equation}
q_c= \begin{cases} 1, & c=\mathcal{C}(L_\text{true}),\\
                    0, & \text{otherwise},\end{cases}
\end{equation}
and $\mathcal{C}(L)$ denotes the class index associated with the true path number $L$. The loss function of the PCD module is then defined as the cross-entropy loss
\begin{equation}
\mathcal{L}_{\mathrm{PCD}}
= -\sum_{c=1}^{C} q_c \log \hat{v}_c.
\end{equation}
By minimizing $\mathcal{L}_{\mathrm{PCD}}$, the network learns to predict an appropriate path-count level for each CSI sample, providing an effective foundation for the subsequent HS-PINNnet introduced in section \ref{sec:method}.

\section{Subchannel-wise Shared Encoding and Parallel Decoding Strategy}
\label{sec: subchannel}
HS-PINNnet reconstructs CSI using multipath parameters. However, for large BS antenna arrays in massive MIMO systems, directly reconstructing the full CSI matrix significantly increases the dimensionality and learning difficulty of network, which issue is particularly important for future XL-MIMO systems. Although the CSI dimension grows with the number of antennas, the underlying propagation environment does not change accordingly. This motivates us to decompose the large-array CSI processing problem into several subchannel reconstruction tasks and propose a subchannel-wise shared encoding and parallel decoding strategy, as shown in Fig. \ref{fig:subchannel}.

Specifically, the CSI matrix in the frequency-spatial domain is divided along the spatial dimension into $N_\mathrm{sub}$ subchannels. According to \eqref{eq:H}, the spatial-domain response of the $l$-th path is determined by
\begin{equation}
\mathbf{a}(\theta_l)=
\left[
1,
e^{j\mu_l},
\ldots,
e^{j(N_\mathrm{t}-1)\mu_l}
\right]^T,
\;
\mu_l=\frac{2\pi d\sin(\theta_l)}{\lambda}.
\end{equation}
For the $s$-th subchannel, let $n_s$ denote its reference antenna index. Since each subchannel has the same number of antennas $N_{\mathrm{s}}$, the local steering vector within each subchannel has the same form:
\begin{equation}
\mathbf{a}_{\mathrm{sub}}(\theta_l)
=
\left[
1,
e^{j\mu_l},
\ldots,
e^{j(N_{\mathrm{s}}-1)\mu_l}
\right]^T.
\end{equation}
Accordingly, the CSI of the $s$-th subchannel is
\begin{equation}
\mathbf{H}_s
=
\sum_{l=1}^{L}
\beta_l e^{j\phi_{l,s}}
\mathbf{b}(\tau_l)\mathbf{a}_{\mathrm{sub}}^H(\theta_l),
\;
\phi_{l,s} = \phi_l + n_s\mu_l .
\end{equation}
This formulation shows that $\beta_l$, $\tau_l$, and $\theta_l$ are shared across subchannels, while $\phi_{l,s}$ varies with the reference antenna index $n_s$.
Based on this physical structure, a shared encoder is recurrently applied to all subchannels to extract compact features in a unified representation space. The resulting features are concatenated and further compressed into the final feedback codeword by an FC layer, aggregating local subchannel representations and global inter-subchannel correlations.

At BS, $N_\mathrm{sub}$ parallel decoders are adopted to reconstruct different subchannels. Each decoder processes the whole codeword, so that the inter-subchannel correlations can be exploited during reconstruction. The outputs of all decoders are finally concatenated along the spatial dimension to recover the full CSI. In this way, the high-dimensional large-array CSI reconstruction is converted into multiple low-dimensional tasks, reducing training difficulty and improving the scalability of the proposed framework for future XL-MIMO systems.

\section{Simulation Results}
\label{sec: results}
\subsection{Dataset and Experimental Setups}


\begin{table}[t]
\centering
\caption{Simulation and Training Parameters}
\label{tab:dataset}
\renewcommand{\arraystretch}{1.2}
\setlength{\tabcolsep}{3pt}
\begin{tabular*}{0.8\linewidth}{@{\extracolsep{\fill}}l l}
\hline
\textbf{Parameter} & \textbf{Setting}\\
\hline
Scenarios & O1, I3 \\ 
Coverage area & $100\,\mathrm{m}\times40\,\mathrm{m}$ (O1) \\
& $10\,\mathrm{m}\times11\,\mathrm{m}$ (I3)\\
Carrier frequency & 3.5 GHz (O1), 2.4 GHz (I3) \\ 
Bandwidth & 50 MHz \\
Subcarriers & $N_\mathrm{c}=64$ \\ 
UE antennas & $N_\mathrm{r}=1$ (also $2,4$ for O1) \\ 
BS antennas& $N_\mathrm{t}=32,256$ \\
Maximum paths & $L=4$ (O1), $L=10$ (I3)  \\ 
Dataset size & 20,000  \\
Epochs & 300 \\
Learning rate & $10^{-4}$ \\
Batch size & 100 \\
\hline
\end{tabular*}
\end{table}

To evaluate the effectiveness of HS-PINNnet, CSI samples are generated using the DeepMIMO platform \cite{deepmimo}, where each sample consists of a channel matrix and the corresponding multipath parameter labels. 
In practical systems, such path-level labels can be obtained from measured channels using conventional high-resolution parameter estimation methods \cite{NOMP_algorithm}.
We consider both outdoor and indoor scenarios, namely the O1 scenario at 3.5 GHz and the I3 scenario at 2.4 GHz. 
BS is equipped with $N_\mathrm{t}=32$ antennas and UE with $N_\mathrm{r}=1$ antenna, while additional configurations with $N_\mathrm{r}=2,4$ (Section~\ref{sec:mUE}) and $N_\mathrm{t}=256$ (Section~\ref{sec:mBS}) are also evaluated for O1. 
The maximum number of paths is set to $L=4$ for O1 and $L=10$ for I3.
Each scenario contains 20,000 samples, split into training, validation, and test sets with a ratio of 8:1:1. HS-PINNnet (both guidance and refinement stages) and the PCD module are trained for 300 epochs using Adam optimizer, with a learning rate of $0.0001$ and a batch size of 100. Detailed scenario and training parameters are summarized in Table~\ref{tab:dataset}.

\subsection{Metrics and Baselines}
We evaluate our model's performance using NMSE and squared generalized cosine similarity (SGCS).
Model complexity is evaluated in floating point operations (FLOPs) and ``FPGA simulation inference latency.'' 
We compare HS-PINNnet with the following baselines: \textbf{TVAL3} \cite{cs}, a CS algorithm for reconstructing the original image from a small number of measurement data. \textbf{CsiNet} \cite{csinet}, the first work combining DL with CSI feedback; \textbf{CsiNet+} \cite{CsiNet+}, which utilizes larger receptive field of convolutional layers than CsiNet to compress and reconstruct CSI; \textbf{CRNet} \cite{crnet}, which adopts a multi-resolution DNN to refine the CsiNet decoder; \textbf{TransNet} \cite{TransNet}, the state-of-the-art model, which introduces a two-layer transformer and applies an attention mechanism in place of traditional convolution operations; \textbf{HS-CsiNet}, which adopts the HS convolutional encoder while utilizes the decoder of CsiNet \cite{csinet}; \textbf{NOMP} \cite{NOMP_algorithm}, which uses an iterative algorithm to estimate multipath parameters from CSI, and then reconstruct the CSI with the iterated parameters; 
and \textbf{Upper Bound of \cite{parameterDL}}, an ideal scheme that assumes the ground-truth parameters are fed back to the BS without any distortion, enabling a clearer comparison with the proposed HS-PINNnet.

\begin{table*}[!t]
\centering
\small 
\setlength{\tabcolsep}{1mm} 
\caption{Comparison of CSI reconstruction accuracy and model complexity across different methods in O1 scenario. NMSE is measured in decibels (dB), SGCS in percent (\%), and FLOPs in millions (M).}
\renewcommand{\arraystretch}{1.1}
\label{tab:results_ratio}
\begin{tabular}{*{16}{c}} 
\toprule
\multirow{2}{*}{Method} &
\multicolumn{3}{c}{$M=4$} & 
\multicolumn{3}{c}{$M=6$} & 
\multicolumn{3}{c}{$M=8$} &
\multicolumn{3}{c}{$M=10$} &
\multicolumn{3}{c}{$M=12$} \\
\cmidrule(lr){2-4} \cmidrule(lr){5-7} \cmidrule(lr){8-10} \cmidrule(lr){11-13}\cmidrule(lr){14-16}
 & NMSE & SGCS  & FLOPs & NMSE & SGCS  &FLOPs & NMSE & SGCS  & FLOPs & NMSE & SGCS & FLOPs& NMSE & SGCS  & FLOPs\\
\cmidrule(lr){1-16}

TVAL3\cite{cs}      & 0.09 & 25.60  &--
            & 0.12 & 25.60 & --
            & 0.16 & 25.60 & --  
            & 0.18 & 28.90 & --
            & 0.20   & 28.90 & --\\
CsiNet\cite{csinet}        & -2.48 & 66.67 &13.88
            & -3.14 & 69.88  &13.95  
            & -3.25 & 69.93 & 14.01 
            & -4.07 & 70.66  &14.08   
            & -4.65 & 75.91  &14.14 \\
CsiNet+\cite{CsiNet+}      & -3.52 & 62.14 &92.16  
            & -4.73 & 68.97   &92.23 
            & -6.23 & 81.99 &92.29 
            & -7.00 & 84.20   &92.36 
            & -7.41 & 84.93  & 92.42\\
CRNet\cite{crnet}       & -3.22 & 61.32  &12.01
            & -4.43 & 75.65  &12.08  
            & -5.84 & 80.95  &12.14  
            & -6.81 & 83.57  &12.21 
            & -6.91 & 84.29  &12.28  \\ 
TransNet\cite{TransNet}    & -5.00 & 70.04  &152.11
            & -8.60 & 87.40 &152.18 
            & -9.90 & 91.33 &152.24  
            & -10.29 & 91.52 &152.31 
            & -12.03   & 94.65 &152.38\\
HS-CsiNet    &-3.92  &61.34   &15.81
            &-6.34  &81.68  &15.87
            &-6.82  &84.45  &15.94  
            &-7.46  &87.08  &16.00 
            &-7.90  &87.67  &16.07 \\
HS-PINNnet   & \textbf{-5.04} & \textbf{70.83}  &\textbf{11.09}
            & \textbf{-9.82} & \textbf{92.49}  &\textbf{11.16}
            & \textbf{-11.61} & \textbf{94.77}  &\textbf{11.22} 
            & \textbf{-12.03} & \textbf{95.06}  &\textbf{11.29}  
            & \textbf{-12.30} & \textbf{95.29} &\textbf{11.35} \\ 
\bottomrule
\end{tabular}

\end{table*}

\begin{figure}
    \centering
    \includegraphics[width=1\linewidth]{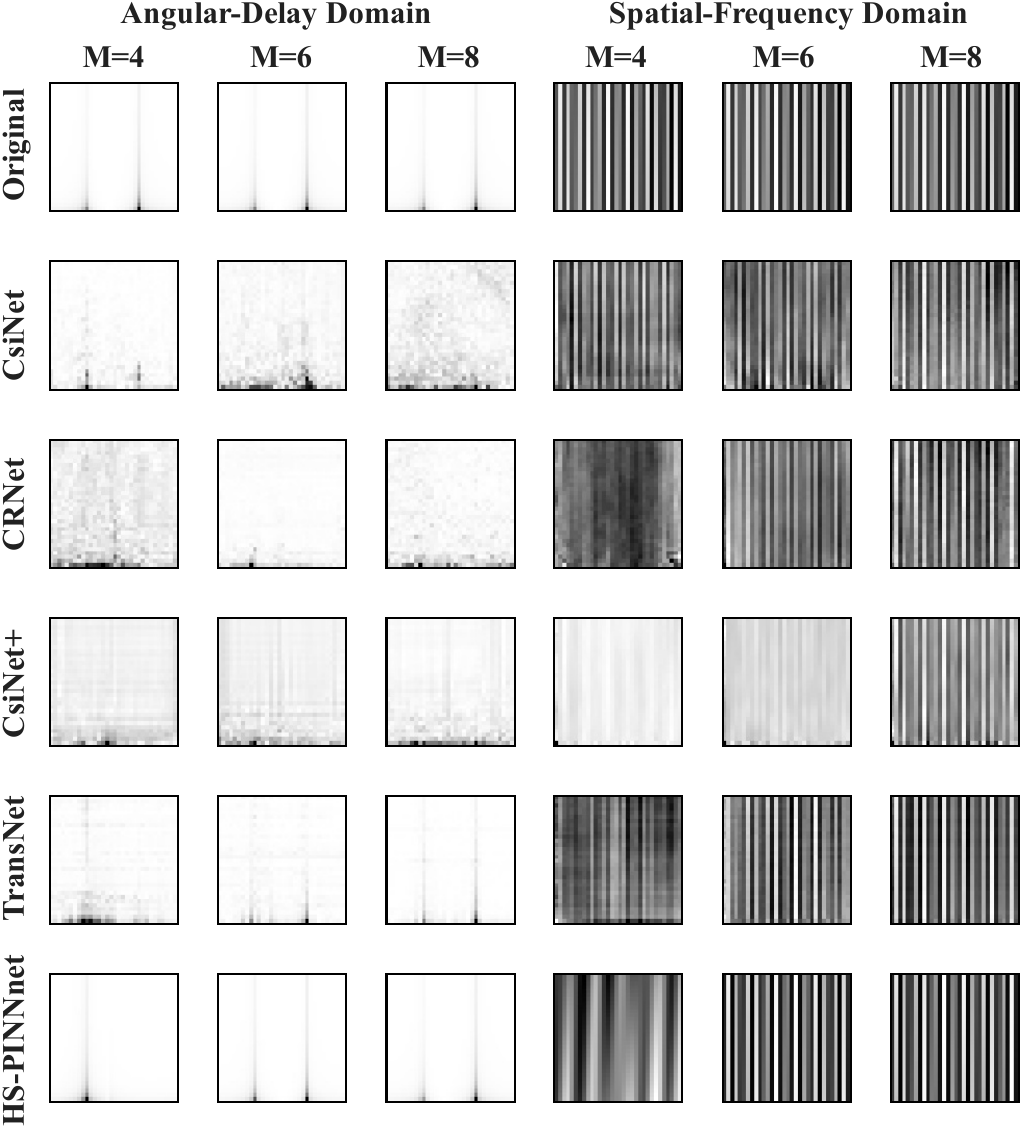}
    \caption{Channel reconstruction visualization of different methods and codeword length in O1.}
    \label{fig:visual}
\end{figure}

\subsection{Performance Under Different Codeword Length}
Table \ref{tab:results_ratio} presents a comparative analysis of HS-PINNnet against baseline models under different compression ratios in O1 scenario. In our simulation setup, we extract 4 multipath components, each characterized by $\beta_l$,  $\phi_l$, $\theta_l$ and $\tau_l$. This results in a maximum of 16 parameters required to reconstruct the CSI. 
However, observations indicate that these parameters often exhibit redundancy (e.g., similar delays across paths). Thus, we set the codeword dimension from $M=4$ to $M=12$ to exploit this redundancy and further compress the information. Table \ref{tab:results_ratio} demonstrates that HS-PINNnet outperforms baseline models across all compression levels, achieving over 5 dB NMSE gain compared to CsiNet. Furthermore, it surpasses TransNet with significantly lower complexity, requiring about 11 million FLOPs versus TransNet’s 152 million FLOPs, which is an approximately 92.8\% reduction.

\begin{figure}
    \centering
    \includegraphics[width=0.95\linewidth]{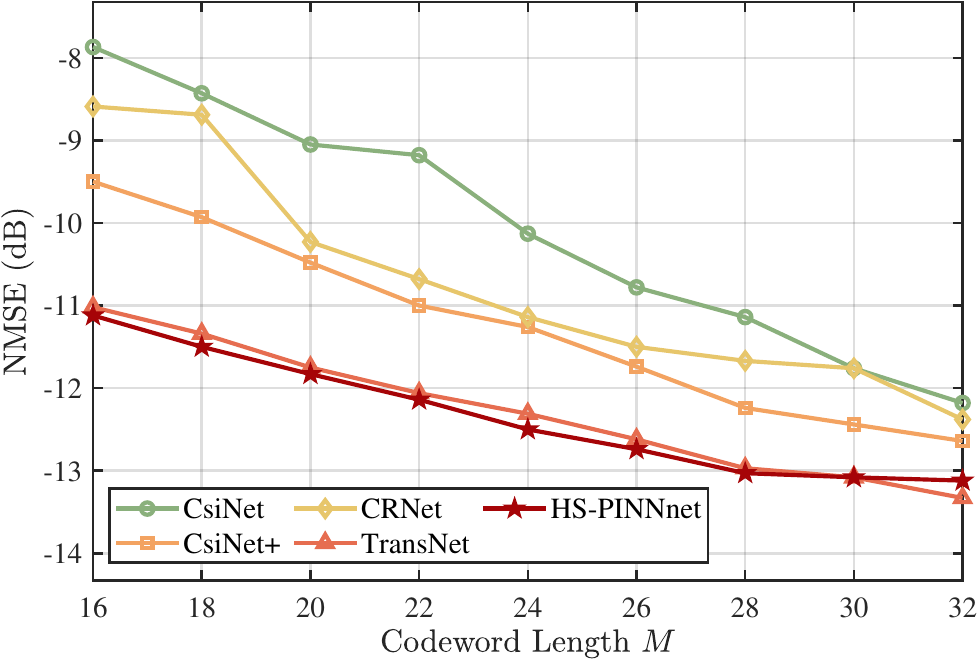}
    \caption{Comparison of CSI reconstruction accuracy of different methods in I3.}
    \label{fig:I3L10}
\end{figure}

The channel reconstruction performance in the indoor I3 scenario is illustrated in Fig. \ref{fig:I3L10}. Compared to the O1 scenario, indoor environment exhibits richer multipath propagation with up to $L=10$ paths. Nevertheless, HS-PINNnet still achieves superior or comparable performance to TransNet with much lower complexity, demonstrating the robustness and scalability of HS-PINNnet across both sparse and dense multipath environments. These advantages, especially at high compression ratios, stem from incorporating the channel model to fully exploit multipath structure, enabling efficient feedback of compact parameter representations with a lightweight architecture.



\subsection{Analysis of the Role of PINN}
To assess the contribution of PINN, we compare HS-PINNnet with HS-CsiNet in Table~\ref{tab:results_ratio}, which adopts the HS convolutional encoder and CNN structure decoder. By reconstructing CSI without incorporating the physical model, HS-CsiNet exhibits an NMSE drop exceeding 3 dB compared to HS-PINNnet. This gap highlights the limitations of purely data-driven approaches, which lack the ability to leverage domain-specific physical knowledge. In contrast, HS-PINNnet integrates a physics-informed decoder to estimate the parameters, effectively exploiting the knowledge of the channel model and achieving superior performance.

Additionally, Fig. \ref{fig:visual} visualizes the reconstructed CSI and only the first 32 subcarriers are shown. The angular-delay domain results show that the original CSI exhibits distinct spreading patterns, each corresponds to an individual propagation path. Conventional DL methods mainly minimize pixel-wise MSE, without explicitly modeling the underlying channel physics. Therefore, they struggle to preserve spreading patterns under high compression ratios and introduce noticeable background noise. In contrast, HS-PINNnet embeds the physical model into reconstruction, enabling physically consistent CSI recovery with clearer spreading structures and noise-free background. Similarly, in the frequency-spatial domain, HS-PINNnet also better preserves the inherent periodic patterns, while conventional methods generate blurred or distorted structures. This comparison validates the advantage of PINN.

\begin{figure}
    \centering
    \includegraphics[width=0.95\linewidth]{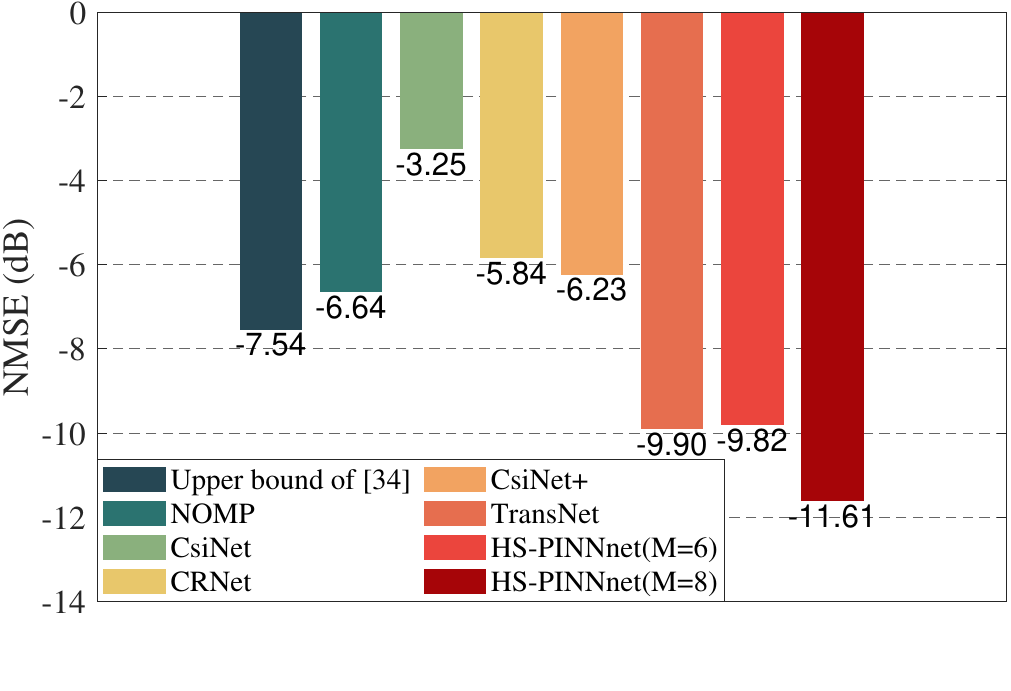}
    \vspace{-3mm}
    \caption{Comparison of CSI reconstruction accuracy when $M=8$.}
    \label{fig:NOMP}
    \vspace{-3mm}
\end{figure}
\subsection{Performance Comparison with Direct Feedback of Multipath Parameters}
As previously discussed, a maximum of 16 parameters is required to reconstruct the CSI, corresponding to four multipath components, each characterized by four parameters. However, due to inherent redundancies in these parameters, such as similar delays across certain paths, the CSI can still be effectively reconstructed even at higher compression ratios. To validate this, performance in O1 scenario with a codeword dimension of $M=8$ is illustrated in Fig. \ref{fig:NOMP}, which corresponds to the parameter count of two propagation paths (i.e., 8 parameters).
In the simulation, the ``Upper bound of \cite{parameterDL}" refers to the ideal scenario where the two strongest propagation paths are perfectly recovered and directly fed back.
In contrast, the NOMP case employs an iterative algorithm to estimate the parameters of the two dominant paths and reconstruct the CSI using these estimates. As shown in Fig. \ref{fig:NOMP}, when $M=8$, HS-PINNnet outperforms both NOMP and upper bound of \cite{parameterDL}, indicating that the codeword generated by HS-PINNnet at $M=8$ not only encodes comprehensive information about the two dominant paths but also retains partial information about the third and fourth paths, thereby enhancing reconstruction accuracy. Remarkably, even at a higher compression ratio with $M=6$, HS-PINNnet still surpasses NOMP, underscoring the encoder’s capability to achieve both effective parameter information extraction and efficient compression. These results highlight the robustness of HS-PINNnet in capturing essential multipath characteristics under strict compression constraints.

\begin{table*}[t]
\centering
\caption{Performance Comparison under Different Numbers of UE Antennas in O1. NMSE is measured in decibels (dB), SGCS in percent (\%), and FLOPs in millions (M). The codeword length is $M=4L$ for $N_r=1$ and $M=5L$ for $Nr=2,4$.}
\label{tab:multi_rx}
\renewcommand{\arraystretch}{1.2}
\begin{tabular*}{0.9\textwidth}{@{\extracolsep{\fill}} c c ccc ccc ccc}
\hline
\multirow{2}{*}{\textbf{Encoded Path}} 
& \multirow{2}{*}{\textbf{Method}} 
& \multicolumn{3}{c}{$\boldsymbol{N_\mathrm{r}=1}$}
& \multicolumn{3}{c}{$\boldsymbol{N_\mathrm{r}=2}$}
& \multicolumn{3}{c}{$\boldsymbol{N_\mathrm{r}=4}$} \\
\cmidrule(lr){3-5} \cmidrule(lr){6-8} \cmidrule(lr){9-11} 
& & \textbf{NMSE} & \textbf{SGCS} & \textbf{FLOPs}
  & \textbf{NMSE} & \textbf{SGCS} & \textbf{FLOPs}
  & \textbf{NMSE} & \textbf{SGCS} & \textbf{FLOPs} \\
\hline
\multirow{2}{*}{$L=1$}
& HS-PINNnet & -5.04 & 70.83 & 11.09 & -6.66 & 79.01 & 11.24 & -7.83 & 83.04 & 11.47 \\
& TransNet   & -5.00 & 70.04 & 152.11 & -6.17 & 77.68 & 317.88 & -6.73 & 80.19 & 686.17 \\
\hline
\multirow{2}{*}{$L=2$}
& HS-PINNnet & -11.61 & 94.77 & 11.22 & -12.17 & 91.24 & 11.32 & -12.59 & 91.07 & 11.55 \\
& TransNet   & -9.90 & 91.33 & 152.24 & -12.06 & 94.59 & 318.05 & -12.11 & 94.90 & 686.50 \\
\hline
\multirow{2}{*}{$L=3$}
& HS-PINNnet & -12.30 & 95.29 & 11.35 & -12.55 & 91.35 & 11.40 & -12.95 & 91.60 & 11.63 \\
& TransNet   & -12.03 & 94.65 & 152.38 & -12.34 & 95.29 & 318.21 & -12.49 & 95.29 & 686.83 \\
\hline
\end{tabular*}
\end{table*}

\subsection{Performance with Multiple UE Antennas}
\label{sec:mUE}
To evaluate scalability, we test HS-PINNnet in the O1 scenario with multiple UE antennas. When $N_\mathrm{r}>1$, an additional angle-of-arrival parameter is required for each path. Therefore, each path is represented by five parameters, and the codeword length is set to $M=5L$. The CSI is stacked along the UE-antenna dimension and fed into the network, while the remaining network architecture is kept unchanged. As shown in Table~\ref{tab:multi_rx}, HS-PINNnet maintains competitive reconstruction performance under different values of $N_\mathrm{r}$ and $L$. More importantly, its FLOPs remains nearly unchanged as $N_\mathrm{r}$ increases. This is because the proposed framework mainly operates on the compact multipath-parameter representation, and only the input(output) dimensions are slightly enlarged. In contrast, TransNet processes CSI in an image-like manner, causing its complexity to increase rapidly with $N_\mathrm{r}$. For example, when $N_\mathrm{r}=4$, TransNet requires more than 686M FLOPs, which is nearly 60 times that of HS-PINNnet. These results demonstrate that HS-PINNnet is more computationally scalable for multi-antenna UE scenarios.

\begin{table}
\centering
\caption{Performance with $N_\mathrm{t}=256$. Other baselines include CsiNet, CsiNet+, CRNet, and TransNet.}
\renewcommand{\arraystretch}{1.3}
\label{tab:large_bs}
\begin{tabular}{c|c|cc}
\toprule
$M$ & Method & NMSE (dB) & SGCS (\%) \\
\hline
\multirow{2}{*}{64} & HS-PINNnet & -5.89 & 81.00 \\
& Other baselines & \multicolumn{2}{c}{Non-convergent} \\
\hline
\multirow{2}{*}{96} & HS-PINNnet & -6.92 & 84.95 \\
& Other baselines & \multicolumn{2}{c}{Non-convergent} \\
\hline
\multirow{2}{*}{128} & HS-PINNnet & -7.86 & 87.17 \\
& Other baselines & \multicolumn{2}{c}{Non-convergent} \\
\bottomrule
\end{tabular}

\end{table}

\subsection{Performance with Larger BS Antenna Arrays}
\label{sec:mBS}
To evaluate the scalability of HS-PINNnet with larger BS antenna arrays, we extend the O1 scenario to $N_\mathrm{t}=256$ and adopt the subchannel-wise shared encoding and parallel decoding strategy, mentioned in Section~\ref{sec: subchannel}, to train the network. $N_\mathrm{sub}$ is set to $8$.
For fair comparison, other baselines are also implemented with the same shared-encoder and parallel-decoder structure.
Since the total parameter dimension is $8\times16=128$, we set $M=64,96,128$. 
As shown in Table~\ref{tab:large_bs}, HS-PINNnet maintains effective CSI reconstruction under the $256$-antenna setting. This is because the proposed framework exploits the shared physical multipath structure across subchannels and compresses redundant parameter information under guiding and refinement training strategy. In contrast, other baselines treat each subchannel as an independent image-like input, making it difficult to exploit such high-dimensional information sharing and rendering their non-convergence at low codeword lengths.

\begin{table}
\centering
{
\caption{Ablation study results under $M=6$.}
\renewcommand{\arraystretch}{1.1}
\label{tab:ablation}
\begin{tabular}{lcccc}
\toprule
Method & NMSE (dB) & SGCS (\%) & FLOPs (M) \\
\midrule
\multicolumn{4}{c}{\textit{\textbf{Encoder Architecture}}} \\
Csi-PINNnet         &-5.65  &83.90  &9.23     \\
Csi-PINNnet (3conv) & -8.28 & 89.62 &15.67   \\

\midrule
\multicolumn{4}{c}{\textit{\textbf{Decoder Architecture}}} \\
Without  main-net &\multicolumn{2}{c}{Non-convergent} & 4.40    \\
Subnet branches 1-1FC  &-4.45  &69.37  &10.11     \\
Subnet branches 1-2FC  &-9.22  &91.45  &10.60     \\
Subnet branches 1-4FC  &-8.77  &91.08  &12.33     \\
Subnet branches 3-3FC  &-8.56  &90.47  &12.21      \\

\midrule
\multicolumn{4}{c}{\textit{\textbf{Training Strategy}}} \\
Without guidance &1.68  &16.43  & 11.16  \\
Without refinement &1.62  &32.46  & 11.16   \\

\midrule
\multicolumn{4}{c}{\textit{\textbf{Proposed Method}}} \\

\textit{\textbf{Ours}} & \textbf{-9.82} & \textbf{92.49} & 11.16 \\
\bottomrule
\end{tabular}}
\end{table}

\subsection{Ablation Experiments}
To further validate our network design, we conduct extensive ablation studies in O1 scenario, summarized in Table \ref{tab:ablation}.

\subsubsection{Encoder Analysis}
We evaluate the impact of the encoder by replacing the hierarchical sensing convolutional mechanism with standard convolutional layers. In the ``Csi-PINNnet'' setting, we substitute the encoder with the one of CsiNet, which adopts a convolutional layer with 2 filters, while the ``Csi-PINNnet (3conv)'' setting employs three convolutional layers with 8, 16, and 2 filters. A performance degradation of more than 1.5 dB is observed in both cases. This confirms the effectiveness of the hierarchical sensing convolutional mechanism, which uses large kernels for global feature extraction and adaptive small kernels for local multipath detail detection.

\subsubsection{Decoder Analysis}
The decoder consists of a main network and a heterogeneous subnet. The main network extracts global features, whose removal causes non-convergence.
Moreover, Table \ref{tab:ablation} compares different FC depths of subnet, where ``$a$-$b$ FC'' denotes $a$ FC layers for shallow branches and $b$ FC layers for deep branches. In the ``1-b FC" cases, the branches for $\beta$ and $\tau$ each employ a single FC layer, as the guiding step shows that one layer suffices to achieve NMSE below -15 dB for both parameters. In contrast, the recovery accuracy of $\phi$ and $\theta$ with a single layer falls short of -10 dB, thus deeper FC branches are adopted. We evaluate $b=1$ to $b=4$ and find that 2 to 4 FC layers accurately recover these angles, with the proposed 3-layer achieving the best performance. In the ``3-3 FC" case, deep branches for $\beta$ and $\tau$ underperform the proposed one due to possible overfitting, further validating the rationality of the subnet.

\subsubsection{Training Strategy Analysis}
The training strategy consists of a guidance and a refinement stage. As shown in Table \ref{tab:ablation}, omitting Step 1, directly using NMSE as the optimization objective, renders HS-PINNnet ineffective. This is because Step 1 is crucial for learning the map between $\mathbf{H}$ and $\mathbf{P}$. Without this step, the network lacks guidance for parameter extraction.
Similarly, skipping the refinement step, directly reconstructing the CSI using the parameters obtained from guidance, also results in poor performance. This is because the exponential and trigonometric operations involved in CSI reconstruction amplify any errors in the estimated parameters, which has been verified in section~\ref{sec:error analysis}. Therefore, Step 2 is essential for refining the parameters to align them with the reconstruction process (\ref{eq:H}).

\begin{figure}
    \centering
    \includegraphics[width=0.95\linewidth]{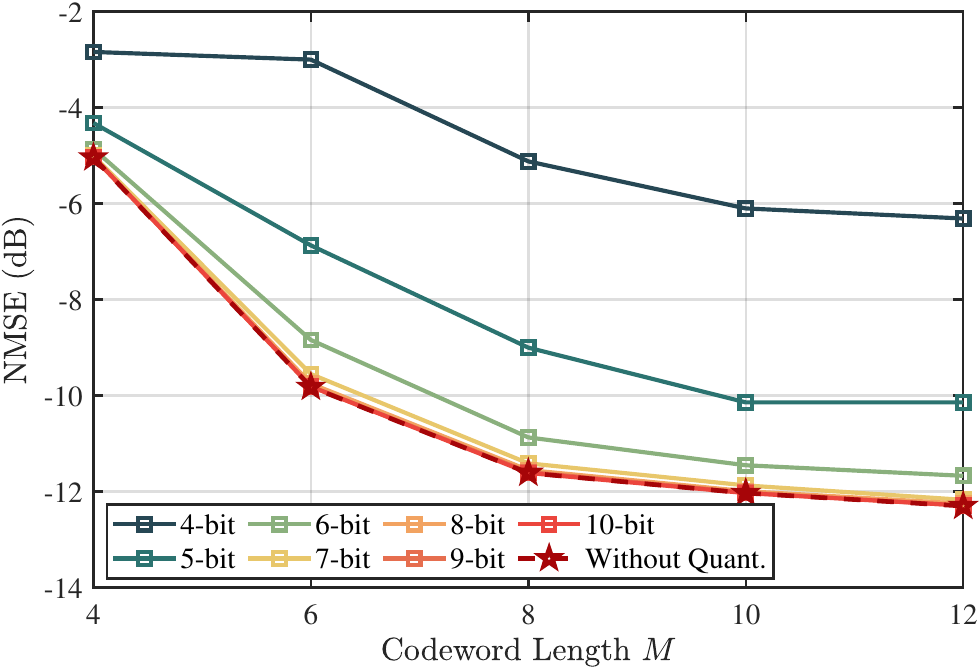}
    \caption{CSI reconstruction performance under different quantization levels.}
    \label{fig:quant}
\end{figure}
\subsection{Codeword Quantization Sensitivity Analysis}
We further investigate the impact of codeword quantization on the performance of the proposed HS-PINNnet in O1 scenario. Since the feedback codeword serves as the compact representation of multipath information, its quantization accuracy directly affects the fidelity of parameter recovery and thus the final channel reconstruction quality. Fig. \ref{fig:quant} shows the channel reconstruction performance under different quantization level, and a clear performance-precision tradeoff can be observed. When the quantization resolution is too low (e.g., 4 bits), the reconstruction performance degrades significantly, indicating that severe information loss occurs during codeword discretization. As the bit increases, the performance improves rapidly and gradually saturates. In particular, 6–7 bit quantization already achieves performance comparable to the unquantized case. Therefore, an appropriate quantization level can be selected to balance feedback overhead and reconstruction accuracy according to system requirements.

\begin{figure*}[t]
    \centering
    \includegraphics[width=0.95\linewidth]{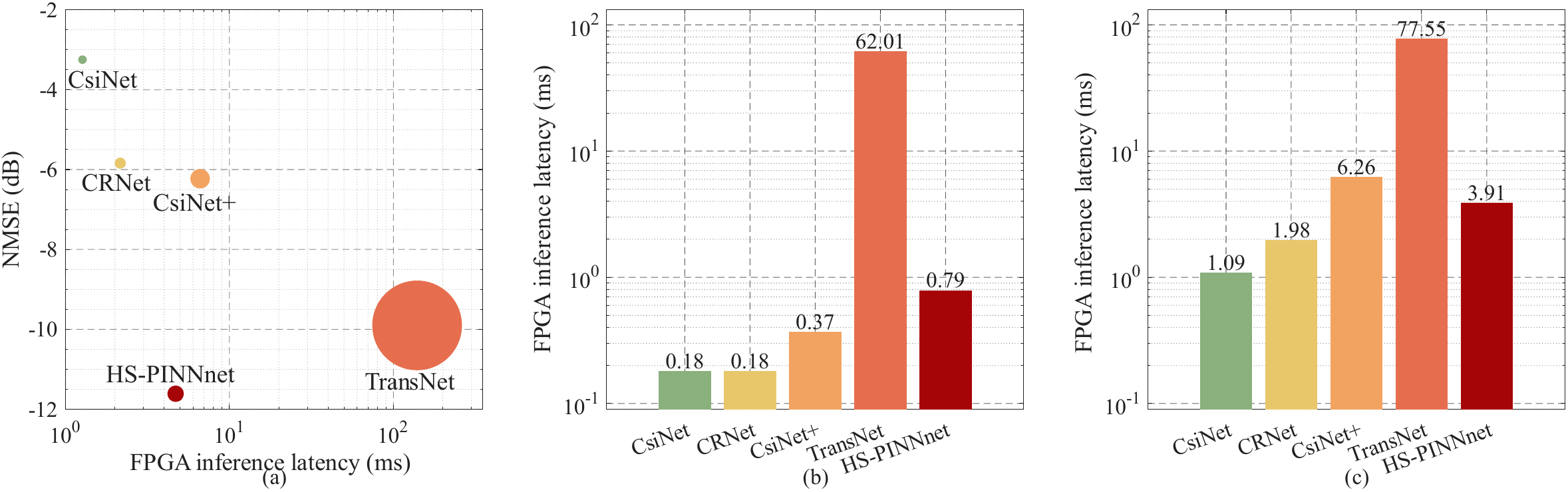}
    \caption{(a) Total FPGA inference latency of different methods. The size of the scatter point is proportional to the inference latency. (b) and (c) is FPGA inference latency of encoders and decoders respectively.}
    \label{fig:FPGA}
\end{figure*}

\subsection{FPGA Complexity Analysis}
Table \ref{tab:results_ratio} reports the FLOPs of different methods. However, FLOPs do not account for data loading, storage, or parallel execution during hardware deployment \cite{FPGA_review}. Therefore, to provide a more realistic estimation of the network's complexity, we employ Deep Learning HDL Toolbox of MATLAB to simulate hardware inference latency. 
The hardware clock frequency is set to 220 MHz, and the estimated latencies for $M=8$ are shown in Fig. \ref{fig:FPGA}(a). Since the toolbox cannot estimate the latency of multihead layer, this portion is excluded from TransNet. It can be observed that although HS-PINNnet achieves the lowest FLOPs in Table \ref{tab:results_ratio}, its latency is slightly higher than that of CsiNet \cite{csinet} and CRNet \cite{crnet}, indicating that FLOPs and latency are not linearly correlated.
Additionally, HS-PINNnet achieves a favorable performance improvement while maintaining relatively low latency, which is two orders of magnitude lower than TransNet. Fig. \ref{fig:FPGA} (b) and (c) further present the encoder and decoder latencies. 
The encoder latency of HS-PINNnet is slightly higher than CsiNet \cite{csinet}, remaining within the same order. This moderate increase brings over 5dB reconstruction gains, thus providing a practical UE-side trade-off between latency and accuracy, while avoiding the unacceptable latency of TransNet.

\begin{table}[t]
\centering
\caption{HS-PINNnet ($M=8$) FPGA performance of each layer when frequency is 220MHz.}
\renewcommand{\arraystretch}{1.1}
\begin{tabular}{lcccc} 
\hline
\textbf{Layer} & \textbf{Latency (ms)} & \textbf{Cycles} & \textbf{FLOPs} & \textbf{$\gamma$} \\
\hline
\multicolumn{5}{c}{\textbf{Encoder}} \\
\hline
Conv0 & 0.07 & 14,748 & 57,344 & 3.89 \\
Conv1\_kernelbranch & 0.33 & 72,438 & 745,472 & 10.29 \\
Conv2\_maskbranch & 0.07 & 14,748 & 57,344 & 3.89 \\
multiple & 0.10 & 21,925 & 16,384 & 0.75 \\
Conv3 & 0.09 & 18,854 & 106,496 & 5.65 \\
FC & 0.13 & 29,018 & 32,768 & 1.12 \\
\hline
\multicolumn{5}{c}{\textbf{Decoder}} \\
\hline
FC & 0.15 & 31,916 & 32,768 & 1.03 \\
Conv4 & 0.05 & 10,642 & 73,728 & 6.93 \\
Conv5 & 0.07 & 16,151 & 294,912 & 18.26 \\
Conv6 & 0.18 & 40,309 & 2,359,296 & 58.53 \\
Conv7 & 0.14 & 31,585 & 589,824 & 18.67 \\
add & 0.05 & 10,993 & 4,096 & 0.37 \\
FC\_branch$\beta$/$\tau$ & 0.07 & 15,771 & 16,384 & 1.04 \\
FC\_branch0 $\phi$/$\theta$ & 1.94 & 426,458 & 524,288 & 1.23 \\
FC\_branch1 $\phi$/$\theta$ & 0.02 & 3,786 & 8,192 & 2.16\\
FC\_branch2 $\phi$/$\theta$ & 0.00 & 289 & 256 & 0.89\\
\hline
\end{tabular}
\label{tab:layer_performance}
\end{table}

Table \ref{tab:layer_performance} shows the latency per layer of HS-PINNnet when $M=8$. ``Cycles'' is the product of the latency and the hardware frequency. The table also includes $ \gamma=\frac{\text{FLOPs}}{\text{Cycles}}$ and a higher ratio indicates faster hardware acceleration.
The results show that convolutional layers exhibit higher $\gamma$ than FC layers, indicating stronger hardware parallelization and acceleration capability. This explains why HS-PINNnet, which contains more FC layers, has slightly longer latency than CsiNet and CRNet. Furthermore, the higher $\gamma$ of ``Conv6'' compared with ``Conv4'' suggests that convolutional layers with more filters can better exploit hardware parallelism.

\begin{table}[t]
\centering
\caption{Performance of the Proposed PCD Module}
\label{tab:PCD_all}
\renewcommand{\arraystretch}{1.2}

\begin{tabular*}{0.8\linewidth}{@{\extracolsep{\fill}}c|cccc}
\hline
\multicolumn{5}{c}{\textbf{Overall Performance}} \\
\hline
\multicolumn{3}{l}{Training Accuracy} & \multicolumn{2}{c}{95.98\%} \\
\multicolumn{3}{l}{Validation Accuracy} & \multicolumn{2}{c}{94.23\%} \\
\multicolumn{3}{l}{Test Accuracy} & \multicolumn{2}{c}{94.74\%} \\
\multicolumn{3}{l}{Relaxed Accuracy} & \multicolumn{2}{c}{96.17\%} \\
\multicolumn{3}{l}{FLOPs} & \multicolumn{2}{c}{0.389M} \\
\hline
\multicolumn{5}{c}{\textbf{Confusion Matrix of Test Set}} \\
\hline
\textbf{True / Predicted} & \textbf{2} & \textbf{4} & \textbf{6} & \textbf{8} \\
\hline
\textbf{2} & 238 & 33 & 0 & 0 \\
\textbf{4} & 31 & 1323 & 2 & 0 \\
\textbf{6} & 0 & 22 & 276 & 10 \\
\textbf{8} & 0 & 0 & 16 & 217 \\
\hline
\end{tabular*}
\end{table}

\subsection{Detection Performance of PCD module}
The performance of the PCD module is evaluated in terms of classification accuracy and computational complexity in the O1 scenario. The confusion matrix in Table \ref{tab:PCD_all} shows that most samples are correctly classified, with only minor confusion between adjacent path-count classes. This behavior is expected, as channels with similar numbers of paths often exhibit comparable propagation characteristics. Quantitatively, the PCD module achieves classification accuracy of 95.98\%, 94.23\%, and 94.74\% on the training, validation, and test set. Furthermore, considering that assigning a smaller path-count sample to a slightly larger class still allows successful recovery, the relaxed accuracy increases to 96.17\%, proving the strong ability of PCD module to capture multipath richness of CSI samples. In terms of complexity, the proposed PCD network requires only 0.389M FLOPs, which is negligible compared to typical CSI feedback models. This demonstrates that the PCD module can be efficiently deployed at the UE side without introducing noticeable computational overhead.

\section{Conclusion}
\label{sec:conclusion}
In this paper, we propose a physics-informed CSI feedback method HS-PINNnet, enabling the network to learn multipath characteristics of the wireless channel. The encoder is designed with a hierarchical sensing convolutional mechanism to efficiently compress the multipath features. The decoder adopts a hetero-architecture to recover each parameter for accurate CSI reconstruction. 
In addition, a PCD module is introduced to adaptively estimate the number of dominant paths contained in CSI samples. A subchannel-wise shared encoding and parallel decoding strategy is designed to improve scalability for applications in future XL-MIMO systems.
Simulations demonstrate that HS-PINNnet advances the performance-complexity trade-off in CSI feedback, outperforming the state-of-the-art method (i.e., TransNet) with simultaneous reductions in computational demands. Specifically, HS-PINNnet achieves a 92.8\% reduction in FLOPs and exhibits two orders of magnitude lower FPGA simulation latency. 
Despite these advances, the current work needs further investigation. First, real-world propagation may involve non-specular scattering, diffuse components, or model mismatch that cannot be fully captured by the idealized parametric multipath model. Second, under high UE mobility, rapid channel variation may affect the stability of path estimation. Future work will extend HS-PINNnet toward more robust modeling under non-ideal propagation and mobility-aware cases.

\renewcommand{\baselinestretch}{1.1}
\bibliographystyle{IEEEtran}
\bibliography{main}

\end{document}